\documentclass[]{aa}  
\usepackage{graphicx}
\usepackage{amsmath}
\usepackage{txfonts}
\usepackage{fourier}
\usepackage{natbib}
\begin{document}

\newcommand{\teff}{T$_{\rm eff}$}
\newcommand{\logg}{$\log${(g)}}
\newcommand{\met}{$[$Fe/H$]$}
\newcommand{\nameofthealgorithm}{\texttt{TOSC}}

%
   \title{TOSC: an algorithm for the tomography of spotted transit chords}


   \author{G. Scandariato\inst{1}{\thanks{e-mail: gas@oact.inaf.it}}, V. Nascimbeni\inst{2,3}, A.F. Lanza\inst{1}, I. Pagano\inst{1}, R. Zanmar-Sanchez\inst{1}, G. Leto\inst{1}}

   \institute{INAF -- Osservatorio Astrofisico di Catania, Via S.Sofia 78, I-95123, Catania, Italy
\and
INAF -- Osservatorio Astronomico di Padova, Vicolo Osservatorio 5, 35122 Padova, Italy
\and
Dip. di Fisica e Astronomia Galileo Galilei – Universit\`a di Padova, Vicolo dell’Osservatorio 2, 35122, Padova, Italy
}


 
  \abstract{Photometric observations of planetary transits may show localized bumps, called transit anomalies, due to the possible crossing of photospheric starspots.}
  {The aim of this work is to analyze the transit anomalies and derive the temperature profile inside the transit belt along the transit direction.}
  {We develop the algorithm \nameofthealgorithm, a tomographic inverse-approach tool which, by means of simple algebra, reconstructs the flux distribution along the transit belt.}
  {We test \nameofthealgorithm\ against some simulated scenarios. We find that \nameofthealgorithm\ provides robust results for light curves with photometric accuracies better than 1~mmag, returning the spot-photosphere temperature contrast with an accuracy better than 100~K. \nameofthealgorithm\ is also robust against the presence of unocculted spots, provided that the apparent planetary radius given by the fit of the transit light curve is used in place of the true radius. The analysis of real data with \nameofthealgorithm\ returns results consistent with previous studies.}
  {}


\authorrunning{G. Scandariato et al.}


   \maketitle
%

\section{Introduction}

During planetary transits, the observed flux of the star decreases because part of the stellar disk is occulted by the planetary disk. If the accuracy of the transit photometry is high enough, it is possible to analyze the fine details of the transit light curve (LC) and characterize the stellar surface. In particular, if the planet transits over stellar spots, then an apparent rebrightening (called \lq\lq transit anomaly\rq\rq) of the star is observed during the spot crossing, as it corresponds to an increase of the flux received from the star because a darker area is occulted.

The analysis of transit anomalies has been theorized by \citet{Schneider2000}. From the observational point of view, the study of \citet{Silva2003} is the first example of how these transit anomalies can be analyzed to obtain information on the morphology and physical parameters of the crossed spots. Since then, many other authors have analyzed the high-accuracy transit photometry of active stars with the aim of modeling the stellar spots \citep[e.g.][]{Pont2007,Rabus2009,Huber2010,Winn2010b,Tregloan2013,Beky2014,Nascimbeni2015,Mancini2015}.

While the analysis of individual transits allows us to model the spots intersected by the planet in those specific epochs, the observations of many transits give us the possibility to study a large number of spots, which can be statistically significant in order to characterize the average photospheric activity \citep{Beky2014}. Furthermore, if the same spots are repeatedly occulted, then the spin-orbit alignment of the system can be constrained \citep{Sanchis2011,Nutzman2011,Desert2011}.

In the case of recurrent anomalies, we can also characterize the time evolution of the spots and the rotation of the star \citep{Silva2008,SilvaValio2011,Sanchis2011b,Beky2014a}. This task is usually accomplished by continuous long-term photometric and spectroscopic monitorings of the stars, with a large requirement of telescope time. Moreover, the results obtained via these methods are affected by several limitations and degeneracies between the parameters of the adopted models \citep{Lanza2016}. Another advantage of transiting systems is then that the evolution of individual spots can be studied using a few consecutive transits.

In recent years, the analysis of transit anomalies has been approached using different methods. For example, some authors fit the transit anomaly with a given analytical model to retrieve the size of the spots and their flux ratio with respect to the photosphere \citep[e.g.][]{Sanchis2011,Nascimbeni2015}. Some other authors have written more sophisticated and computationally expensive codes, which fit the transit light curves by means of Monte Carlo algorithms \citep[e.g.][]{Tregloan2013,Beky2014}. In both cases, some assumptions on the shape and distribution of the spots over the stellar disk are needed.

In this paper we present the Tomography Of Spotted transit Chords (\nameofthealgorithm) code, a new and fast algorithm which analyzes the transit anomalies and reconstructs the flux distribution inside the area of the stellar surface intersected by the transiting planet, namely the \lq\lq transit chord\rq\rq\footnote{The term \lq\lq chord\rq\rq\ is used improperly, as the chord is by definition a straight segment with endpoints on a circle. From a pure geometrical point of view, it would be more appropriate to use the term \lq\lq belt\rq\rq. Nonetheless, in exoplanet science the two terms are used interchangeably. In this paper, we will prefer the notation \lq\lq chord\rq\rq\ as most widely used by the community. We will use the term \lq\lq belt\rq\rq\ only when it is required by the bi-dimensionality of the tomography.}. The main advantage of our code with respect to previous approaches is that it only needs some assumptions on the geometry of the planetary system and on the spectrum of the star. In other words, \nameofthealgorithm\ does not assume any a priori parameter of the photospheric spots occulted by the planet (e.g. shape, size, temperature, etc\dots). In Sect.~\ref{sec:model} we describe the geometrical model and the inversion algorithm which returns the tomography of the transit chord. In Sect.~\ref{sec:simulations} we test \nameofthealgorithm\ against a few simulated transits, in order to check its robustness and its response to photometric accuracy and spot parameters. Finally, in Sect.~\ref{sec:comparison} we compare our code with other models for transit anomalies, applying \nameofthealgorithm\ to some transit LCs which have already been analyzed in literature.
 
\section{The model}\label{sec:model}

The out-of-transit stellar flux $F_{oot}$ depends on the effective temperature of the star and on the presence of active regions on the stellar surface. Active regions evolve with time and rotate with the star. $F_{oot}$ is thus by principle a function of time. Nonetheless, both effects (evolution and rotation of active regions) generally have timescales much longer than the typical duration of a planetary transit (i.e.\ a few hours). It is thus a common practice to fit a low-order polynomial to the out-of-transit data, and to normalize the observed fluxes by this polynomial, in order to remove any long-term trend in the observed LC. After normalization, the dependence of $F_{oot}$ on time can thus be dropped. In the following discussion, all the fluxes are normalized by the low-order polynomial fit of the out-of-transit LC.

During the transit, the flux $F(t)$ received from the stellar disk is given by:
\begin{equation}
F(t)=F_{oot}-F_{occ}(t),\label{eq:fluxAbs}
\end{equation}
where $F_{occ}(t)$ is the occulted stellar flux at any time $t$. The dependence on time is given by the fact that $F_{occ}(t)$ depends on the limb darkening \citep[LD, see e.g.][]{Winn2010} profile. Moreover, the light curve may have anomalies in case the planet occults active regions, either in the form of cool spots or photospheric bright faculae, showing bumps or dips correspondingly \citep[e.g.][]{Silva2003,Ballerini2012}.

$F(t)$ is the observable that we want to convert into the spatially-resolved brightness profile of the transit chord. To this purpose, in the following sections we describe how we model the stellar disk (Sect.~\ref{sec:stellarDisc}), the transit chord (Sect.~\ref{sec:transitChord}) and the transit LC (Sect.~\ref{sec:algorithm}).

\subsection{The stellar disk}\label{sec:stellarDisc}

If we neglect the effects of surface brightness inhomogeneities and of their rotation over timescales of a few hours typical of the transits, we can assume that the flux emerging from the stellar disk is constant, and is given by the surface integral of the specific intensity $I$ over the disk.

The stellar disk observed from Earth is not uniformly bright due to the LD effect. We assume a quadratic profile \citep{Claret2004}, which is generally adopted in fitting transit LCs. This profile is usually parametrized as:
\begin{equation}
\frac{I(\mu)}{I_0}=1-u_1(1-\mu)-u_2(1-\mu)^2,\label{eq:ld}
\end{equation}
where $I_0$ is the specific intensity at the center of the disk, $\mu=\cos\gamma$, $\gamma$ is the angle between the line of sight and the normal to the stellar surface, and $u_1$ and $u_2$ are the limb darkening coefficients, which are part of the output of the transit LC fitting or are derived from a model atmosphere.

To compute the broadband intensity $I_0$ at the center of the stellar disk, we first convolve the intensity given by the BT-Settl synthetic spectral library by \citet{Baraffe2015} with the response function of the instrumental setup used to collect the LC. Then, the intensity $I_0$ is computed interpolating the temperature grid of the model at the effective temperature \teff\ of the star.

The flux $F_{oot}$ emerging from the stellar disk is thus given by the integral of $I(\mu)$ over the stellar disk \citep[see Eq.~3 in][]{Lanza2016}:
\begin{equation}
F_{oot}=2\pi\ R_*^2\int_0^1 I(\mu)\mu d\mu.\label{eq:fluxIntegral}
\end{equation}

Substituting Eq.~\ref{eq:ld} into Eq.~\ref{eq:fluxIntegral} and integrating we obtain:
\begin{equation}
F_{oot}=\pi R_*^2I_0\frac{6-2u_1-u_2}{6},
\end{equation}
where $R_*$ denotes the stellar radius.

\subsection{The transit chord}\label{sec:transitChord}

Assuming that the shift of the planet on the stellar disk during the photometric exposure is negligible compared to the planetary radius, the stellar flux $F_{occ}(t)$ occulted by the planet at the time $t$ during the transit is given by the integral of the stellar flux over the planetary disk in the sky-projected plane. We can thus assume that each photometric point taken during the transit is representative of an area on the projected stellar disk centered on the planet and having the same planetary radius. Hence, as the planet moves over the stellar disk, it scans the transit chord. The transit LC is thus, from this point of view, the observable carrying information on the flux distribution along the transit chord.

In particular, the U-shaped transit LC is due to the LD, while anomalies (bumps and/or dips) with respect to the best-fit transit model are due to active regions on the stellar surface. In both cases, the inference on LD and/or stellar activity is possible because the planetary radius is smaller than the length of the transit chord (the planetary radius is of the order of 10\% or less of the stellar radius), thus the movement along the transit chord provides a spatially resolved scan of the chord itself. Roughly speaking, the spatial resolution of the scan is of the same order of the shift of the planet along the chord between two adjacent photometric points if the noise can be neglected.

To reconstruct the flux distribution along the transit chord, we refer the stellar disk to a Cartesian reference frame centered at its center. We assume the transit chord to be in the horizontal direction, at the ordinate $y=b\cdot R_*$, where $R_*$ is the stellar radius and $b$ the impact parameter in units of the stellar radius.

\begin{figure*}
\centering
\includegraphics[width=.66\linewidth]{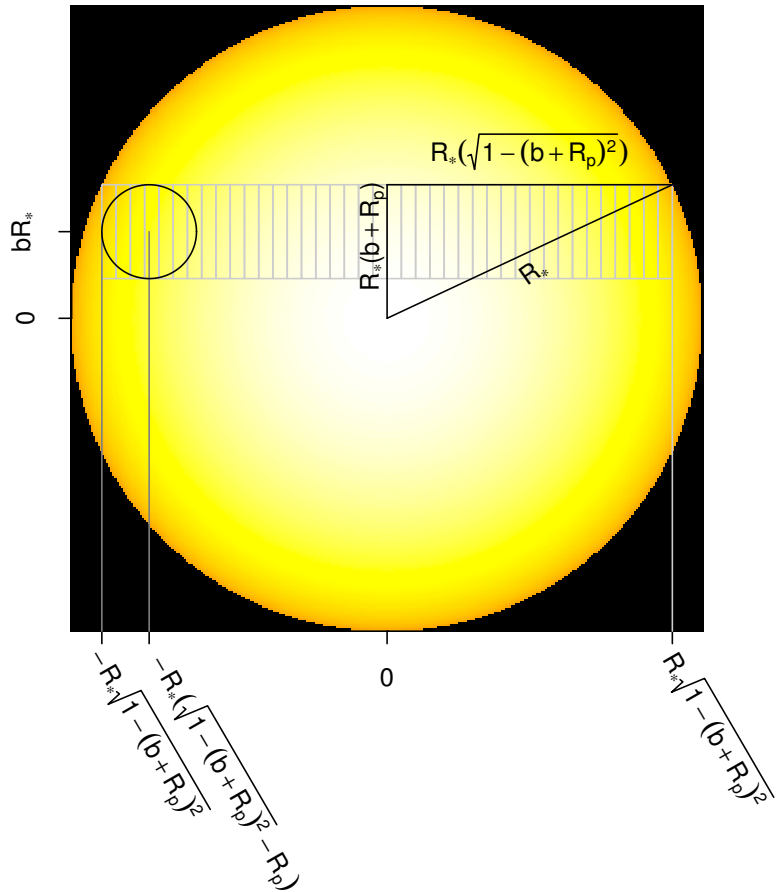}
\caption{Schematic representation of the geometry discussed in the text. The circle represents the transiting planet, while the grid rectangles show the cells used in the chord reconstruction.}\label{fig:geometry}
\end{figure*}

For the sake of simplicity, in our model the disk of the planet is assumed to be entirely inside the disk of the star. In other words, we do not model the ingress and the egress. We thus define a grid inside the transit chord made up of adjacent equal rectangular cells\footnote{If the orbit inclination is not 90$^\circ$, the planetary track across the stellar disk is actually curved. The most tilted configuration occurs when the projected planetary orbit is an ellipse of semi minor axis equal to the stellar radius. In this case, the maximum displacement along the vertical direction of the planet during the transit is $R_*\sqrt{1-(R_*/a)^2}$, where $a$ is the semi major axis. For a giant planet ($R_p\sim0.1$) in a tight orbit ($R_*/a\sim0.1$), the maximum displacement turns out to be a few percents of the planetary radius. Given this small effect, and the difficulty in taking into account warped cells inside the curved transit chord, for the sake of simplicity we approximate the transit chord as a rectilinear segment.}, each one with height equal to the diameter of the planet and width $\Delta x$ depending on the number $N$ of cells (Fig.~\ref{fig:geometry}). The length of the grid equals the length of the top side of the chord in Fig.~\ref{fig:geometry}, such that the first and last grid cells are completely contained into the projected stellar disk. Using Pythagoras's theorem, the total length $l_{chord}$ of the reconstructed chord is:
\begin{equation}
l_{chord}=2R_*\sqrt{1-\left(R_p+b\right)^2},\nonumber
\end{equation}
and the width of each cell is thus given by
\begin{equation}
\Delta x=\frac{l_{chord}}{N}=\frac{2R_*\sqrt{1-\left(R_p+b\right)^2}}{N},\label{eq:deltax}
\end{equation}
where $R_p$ is the planetary radius in units of the stellar radius.

For the sake of simplicity, we assume that the intensity $I_i$ emerging from the $i$-th cell in the chord is uniform and equal to the intensity at the center of the cell:
\begin{equation}
I_i=I\left(\mu(x_i,b\cdot R_*)\right),\label{eq:approximation}
\end{equation}
(i.e.\ we neglect the dependence of $I$ on $\mu$ inside the cell, see Eq.~\ref{eq:ld}). The corresponding emerging flux is thus:
\begin{align}
F_i=&\iint I(x,y) dx dy\simeq I\left(\mu(x_i,b\cdot R_*)\right)\iint dx dy=\nonumber\\
=&I\left(\mu(x_i,b\cdot R_*)\right)\cdot 2R_pR_*\Delta x.
\end{align}
where the double integral is extended over the surface of the $i$-th cell.

\subsection{The reconstruction algorithm}\label{sec:algorithm}

The flux dimming $F_{occ}(t)$ at any time $t$ during the transit is the sum of the fluxes $F_i$ of the cells multiplied by the corresponding overlap fractional area between the cells and the planetary disk. With these assumptions, we obtain:
\begin{equation}
F_{occ}(t)=\sum_{i=1}^{N}w_i(t)\cdot F_i,\label{eq:occultation}
\end{equation}
where the weights $w_i(t)$ are the overlap fractional areas, and thus run from 0 to 1.

\begin{figure}
\centering
\includegraphics[width=\linewidth]{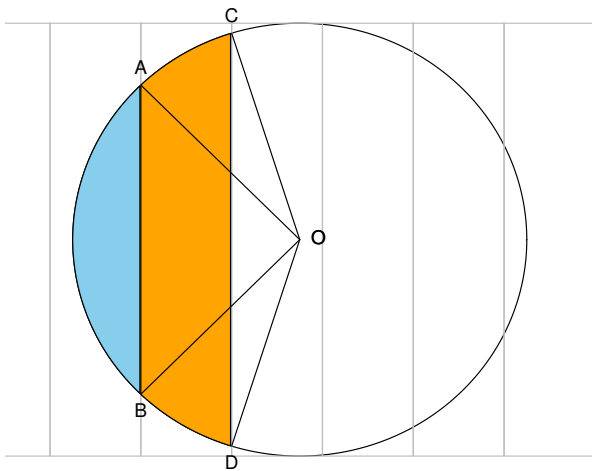}
\caption{Graphical representation of the computation of the weights discussed in the text. The circle is the planetary disk centered in $O$, while the grey grid shows the cells drawn on the stellar disk partially covered by the planetary disk.}\label{fig:intersection}
\end{figure}

The weights $w_i(t)$ are computed by geometric considerations (Fig.~\ref{fig:intersection}). From left to right, the first non-zero weight is computed as the area (in cyan in Fig.~\ref{fig:intersection}) of the circular segment bounded by the arc $\widearc{AB}$. The second weight (i.e.\ the orange area in Fig.~\ref{fig:intersection}) is then computed as the area of the circular segment bounded by the arc $\widearc{CD}$ (the cyan and orange circular segment) minus the area of the circular segment bounded by  $\widearc{AB}$, and so forth. 

The $N$ unknowns $F_i$ can be retrieved solving the linear system:
\begin{equation}
\begin{bmatrix} w_1(t_1) & \cdots & w_N(t_1) \\ \vdots & \ddots & \vdots \\ w_1(t_M) & \cdots & w_N(t_M) \end{bmatrix}
\cdot
\begin{bmatrix} F_1 \\ \vdots \\ F_N\end{bmatrix}
=
\begin{bmatrix} F_{occ}(t_1) \\ \vdots \\ F_{occ}(t_M)\end{bmatrix}\label{eq:system}
\end{equation}
where $t_1\dots t_M$ are the times when the $M$ photometric points were collected.

In principle, there is no restriction on the maximum number of cells. Nonetheless, large values of $N$ are discouraged for two reasons. First of all, if $N>M$ then the system in Eq.~\ref{eq:system} is overdetermined and the data are \lq\lq overfitted\rq\rq. This corresponds to extracting from the data an amount of information larger than the information carried by the data themselves.

The second reason is given by the Nyquist sampling theorem. Assuming that the transit is evenly sampled at positions separated by the distance $\Delta x_0$ on the stellar disk, the planet samples the transit chord with the spatial frequency $f_0=\tfrac{1}{\Delta x_0}$. The corresponding Nyquist frequency is $f=f_0/2$, which is the maximum frequency that allows the reconstruction of the transit chord without aliases. The Nyquist theorem thus suggests to divide the chord into cells whose minimum width is $\Delta x_{min}=2\Delta x_0$. Plugging this into Eq.~\ref{eq:deltax}, the requirement on the spatial frequency translates into $N<M/2$.

We let the user of \nameofthealgorithm\ decide the optimal $N$ for his purposes, cautioning that in our tests values of $N$ larger than $M/2$ have proven to introduce aliases in the reconstruction.

Noise in the data, numerical precision and the fact that the system is overdetermined if $N<M$ generally lead to the lack of an exact solution for Eq.~\ref{eq:system}. For these reasons, we solve Eq.~\ref{eq:system} via a $\chi^2$ minimization procedure rather than by using canonical linear methods, aiming at the vector $\widetilde{x}$ which best approximates Eq.~\ref{eq:system} via the condition
\begin{equation}
\widetilde{x}=\min_x\left(\lVert A\cdot x-F_{occ} \rVert^2\right),\label{eq:solution}
\end{equation}
where $A$ is the $M\times N$ matrix of the coefficients in Eq.~\ref{eq:system}
\setcounter{MaxMatrixCols}{20}
\begin{equation}
A=\begin{bmatrix}
w_1(t_1) & & & & & \dots & & & & & w_N(t_1)\\
\vdots   & & & & &  & & & & & \vdots\\
w_1(t_M) & & & & & \dots & & & & & w_N(t_M)\\
\end{bmatrix};\label{eq:Amatrix}
\end{equation}
and
\begin{equation}
F_{occ}=\begin{bmatrix}
F_{occ}(t_1)\\
\vdots\\
F_{occ}(t_M)\\
\end{bmatrix}.\label{eq:focc}
\end{equation}

To improve the capability of \nameofthealgorithm\ to converge to a physically acceptable solution, in \nameofthealgorithm\ there is the possibility to constrain the fitted cell flux\footnote{To implement the minimization algorithm with constraints inside \nameofthealgorithm, we use the function \texttt{lsei} of the \texttt{limSolve} package v.1.5.5.1 \citep{Soetaert2009} for the \texttt{R} environment \citep{R2016}, which uses the singular value decomposition method for matrix inversion \citep{Press1992}.}.  In the assumption that the transit LCs are not affected by bright faculae, the maximum flux expected from the cells cannot exceed the photospheric flux. The user can thus constrain the maximum flux $F_{i,max}$ corresponding to $T_{\rm eff,max}$=\teff+$n\sigma_{\rm T_{eff}}$, where $\sigma_{\rm T_{eff}}$ is the measurement uncertainty on the stellar \teff, while $n$ is a free parameter set by the user. Setting $n=0$, the maximum allowed temperature equals the nominal stellar \teff; setting $n$ to a very large number corresponds to putting no constraint on $T_{\rm eff,max}$. Moreover, the user can also constrain the minimum flux $F_{i,min}$ emerging from the grid cells by setting the minimum temperature $T_{\rm eff,min}$ for the reconstructed transit chord. To this purpose, the study of \citet{Berdyugina2005} and \citet{Andersen2015} provide some useful indications on the spots' temperature contrast as a function of the stellar effective temperature.

Mathematically speaking, solving Eq.~\ref{eq:solution} is an ill-posed problem. Hence, we regularize the inversion problem using Tikhonov's method, i.e. we insert in Eq.~\ref{eq:solution} an additive term such that the best solution $\widetilde{x}$ is found as:
\begin{equation}
\widetilde{x}=\min_x\left(\lVert A\cdot x-F_{occ} \rVert^2+ \lambda\lVert T\cdot x\rVert^2\right),\label{eq:solutionComplete}
\end{equation}
where $\lambda$ is a non-negative Lagrangian multiplier and $T$ is, in the Tikhonov regularization framework, an ad-hoc matrix which gives preference to solutions with acceptable characteristics. We give preference to smooth solutions of Eq.~\ref{eq:solutionComplete} by minimizing the differences between neighboring elements of $x$, with the aim of providing solutions with minimum jumps between neighboring cells along the transit chord\footnote{The Tikhonov regularization is suited for the cases where some degree of correlation between neighboring points, as in the case of extended spotted areas inside the transit chord. An alternative is the Maximum Entropy regularization, which conversely minimizes the correlation between different parts of the reconstructed image \citep{Piskunov1990}, and is thus less suited for the aim of \nameofthealgorithm.}.


\begin{equation}
T=\begin{bmatrix}
1 & -1 & 0 & 0 & \dots & 0 & 0 \\
0 & 1 & -1 & 0 & \dots & 0 & 0 \\
\vdots & &&&&&\vdots\\
0 &  &  &  & \dots & 1 & -1 \\
\end{bmatrix}.\label{eq:Tmatrix}
\end{equation}

The constraints and the regularization through the Lagrangian multiplier $\lambda$ are useful to reduce the propagation of errors and numerical instabilities introduced by approximations and noisy data. A posteriori we also find that the regularization in the regression algorithm is able to reduce the autocorrelation in the residuals introduced by numerical approximations in the code (see Sect.~\ref{sec:simulations}). The side effect of the regularization is to smooth the solution of the reconstruction problem, lowering the spatial resolution of the reconstruction. In particular, at large $\lambda$ the solution of Eq.~\ref{eq:solutionComplete} is over-smoothed, leading again to autocorrelated first neighbors. We will further discuss the impact of $\lambda$ on the solution in Sect.\ref{sec:hatp11}, where we apply \nameofthealgorithm\ to a clear transit feature detected in real data.

In \nameofthealgorithm\ we implement an automatic search of the optimal multiplier $\lambda$. This search consists in increasing iteratively the value of $\lambda$ until the absolute value of the autocorrelation between first neighbors decreases below the threshold $1.96/\sqrt{N}$. This threshold corresponds to the 95\% confidence interval for the hypothesis that residuals are randomly distributed around zero \citep{Chatfield1980}. We remark that this is an automatic optimization criterion, which can be overridden by the user in case of unsatisfactory results.

Once the optimal $\lambda$ multiplier is fixed, \nameofthealgorithm\ computes the array of fluxes $\widetilde{x}$ in Eq.~\ref{eq:solutionComplete}, which is then converted into temperatures by means of the grid of fluxes predicted by the synthetic spectral models (Sect.~\ref{sec:stellarDisc}). The temperature contrast $\Delta T$ discussed in the following sections is the difference between the reconstructed and the effective temperature, and is thus negative in the presence of cool spots.

\section{Simulated transits}\label{sec:simulations}
To test \nameofthealgorithm, we check its robustness and accuracy in reconstructing a simulated transit chord. In the following subsections, we simulate a planetary system with the parameters of the planet system hosted by HAT-P-11 (listed in Table~\ref{tab:parameters}), and we analyze the output of the reconstruction algorithm when different configurations of spots are simulated.

\begin{table}
\begin{center}
\caption{Parameters of the planetary system HAT-P-11 used in this paper \citep{Bakos2010}.}\label{tab:parameters}
\begin{tabular}{lr}
\hline\hline
\teff & 4780 K\\
$R_p$ & 0.05862 $R_*$\\
$a$\tablefootmark{a} & 0.0530 AU\\
$i$\tablefootmark{b} & 88.50$^\circ$\\
$e$\tablefootmark{c} & 0.198\\
$\omega$\tablefootmark{d} & 355.2$^\circ$\\
$u_{1}$,$u_{2}$ \tablefootmark{e,f} & 0.599, 0.073\\
\hline
\end{tabular}
\tablefoot{
	\tablefoottext{a}{Orbital semi-major axis.}
	\tablefoottext{b}{Orbital inclination.}
	\tablefoottext{c}{Orbital eccentricity.}
	\tablefoottext{d}{Argument of periastron.}
	\tablefoottext{e}{Quadratic limb darkening coefficients.}
	\tablefoottext{f}{From \citet{Sanchis2011}.}
}
\end{center}
\end{table}

\subsection{Transit curve simulation}\label{sec:transitSimulation}
There are several codes capable to simulate the LCs of planetary transits. These codes generally simulate a clean stellar disk, with no signature of stellar activity. Since in this work we are interested in analyzing the effect of photospheric spots on transit LCs, we write a simple code to simulate the planetary transits over spotted photospheres, to provide to \nameofthealgorithm\ some test cases. In brief, this code pixelates the stellar disk along two perpendicular directions, and for each pixel the emerging flux is computed convolving the photometric passband with the same limb-darkened stellar spectrum used in \nameofthealgorithm. The number of pixels along each direction is chosen one order of magnitude larger than the number of the simulated photometric points (see Sect.~\ref{sec:spotfree}). A posteriori, we find that increasing the number of pixels does not have any effect on the reconstruction of the transit chord.

Then, the flux dimming for any position of the planetary disk over the stellar disk is computed. The integration of the occulted stellar flux is performed using the IDL routine \texttt{aper.pro}, which computes aperture photometry with the capability of counting subpixels. A posteriori, we find that this feature provides a major improvement in the computation of the LC, which in turn allows a more precise reconstruction of the transit chord.

\subsection{Transit over a spot-free stellar disk}\label{sec:spotfree}
The simplest case to test \nameofthealgorithm\ is the transit of the planet over a spot-free stellar disk. To this purpose, we simulate the transit of the planet uniformly sampling the transit chord with 100 points between $-R_*\left(\sqrt{1-\left(b+R_p\right)^2}-R_p\right)$ and $R_*\left(\sqrt{1-\left(b+R_p\right)^2}-R_p\right)$ (see Fig.~\ref{fig:geometry}). This is the typical number of photometric points in transit LCs provided by dedicated space-borne observatories. Following the Nyquist sampling theorem (Sect.~\ref{sec:algorithm}), this gives a number of cells equal to 50.

First of all, we neglect the LD effect and do not include measurement errors in the simulations. In the absence of LD the occulted flux is a constant fraction $R_p^2$ of the stellar flux between the second and third contact. Nonetheless, we use the transit simulator discussed in Sect.~\ref{sec:transitSimulation} to check the inaccuracies introduced by numerical approximations in the transit simulation.

\begin{figure}
\centering
\includegraphics[width=\linewidth]{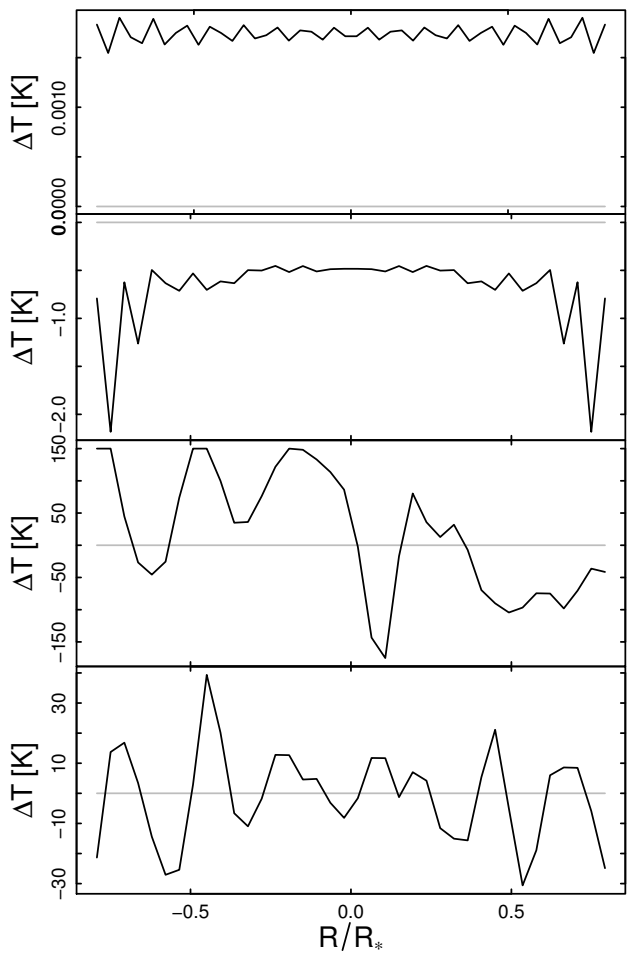}
\caption{Examples of reconstruction of the thermal profile of the transit chord for the case of an unspotted stellar disk. The orbital parameters used in the simulation are listed in Table~\ref{tab:parameters}. From top to bottom, we show the case of (i) a stellar disk with no LD, noise-free LC and $\lambda=0$; (ii) a stellar disk with LD, noise-free LC and $\lambda=0$; (iii) a stellar disk with LD and LC with a photometric precision of 1~mmag and $\lambda=23$; (iv) a stellar disk with LD and LC with a photometric precision of 0.1~mmag and $\lambda=2.9$.}\label{fig:unspottedFits}
\end{figure}

We run \nameofthealgorithm\ on the simulated LC, setting $\lambda$=0 for the regularization discussed in Sect.~\ref{sec:algorithm}. We find that \nameofthealgorithm\ is not capable to exactly return $\Delta T$=0 as expected. The best fit is systematically $\sim10^{-3}$~K off the true value, and it shows a regular autocorrelated pattern with amplitude of the order of $\sim10^{-4}$~K (Fig.~\ref{fig:unspottedFits}, top panel). These features are due to the propagation of numerical approximations both in the LC simulation and in the transit chord reconstruction. We remark that these inaccuracies are negligible compared with the uncertainty usually associated with the measurement of stellar effective temperatures ($\lesssim$100~K), and are thus irrelevant in any practical purpose.

We then simulate the more realistic case of a planetary transit over a limb-darkened stellar disk, using the quadratic LD law with the coefficients in Table~\ref{tab:parameters}. In this case, the unregularized algorithm is able to reconstruct the transit chord temperature profile with an uncertainty of $\lesssim$1~K (see Fig.~\ref{fig:unspottedFits}, second panel from top). We thus find that the presence of LD increases the offset of the best fit solution with respect to the true value. This is due to the fact that the flux assigned to each grid cell is an approximation of the true one (see Eq.~\ref{eq:approximation}).

Finally, we add some randomly-generated noise to the transit LC. First of all we simulate a 1~mmag photometric noise, typical of ground-based high-precision photometric observations. A posteriori, we find that the output of the regression is heavily affected by the propagation of the photometric uncertainties in the matrix inversion such that, in the most extreme cases, the automatic optimization of $\lambda$ does not converge to a smooth solution. When it does, we find that the $\lambda$ multiplier is of the order of $\lesssim$10 and the accuracy in the reconstruction of the transit chord is $\sigma_T\sim$100~K (Fig.~\ref{fig:unspottedFits}, second panel from bottom).


If we decrease the photometric noise down to 0.1~mmag, similar to Kepler's precision \citep{Borucki2010} and to what is expected from future photometers \citep[e.g.\ CHEOPS,][]{Scandariato2016}, we find that the convergence of \nameofthealgorithm\ is more robust: the accuracy in the reconstruction is $\sigma_T\lesssim$30~K (Fig.~\ref{fig:unspottedFits}, bottom panel).

\begin{figure}
\centering
\includegraphics[width=\linewidth]{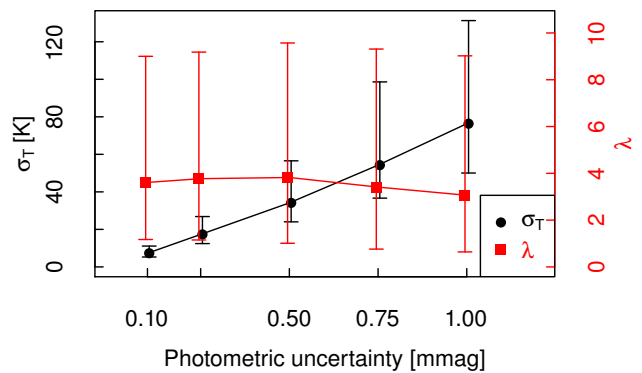}
\caption{$\lambda$ (red squares) and $\sigma_T$ (black circles) as functions of the photometric uncertainty of the LC.}\label{fig:noiseramp}
\end{figure}

To better understand how $\lambda$ and $\sigma_T$ depend on the photometric uncertainty, we run the following test. We simulate the five different values of the photometric uncertainty, from 0.1mmag to 1 mmag, shown in Fig.~\ref{fig:noiseramp}. For each of these values, we generate 10$^5$ random samples of photometric noise to add to the model LC. For each noisy LC we run \nameofthealgorithm\ and we store the corresponding $\lambda$ and $\sigma_T$. For each value of the photometric noise we can thus compute the median $\lambda$ and $\sigma_T$, which are plotted in Fig.~\ref{fig:noiseramp}. We find that $\lambda$ is generally of the order of a few units or less and does not depend on the photometric accuracy, while $\sigma_T$ increases with the photometric uncertainty of the LC, consistently with the test cases discussed above.

\subsection{Transit over a spotted stellar disk}\label{sec:spottedStellarDisk}

Photospheric spots modify the transit LC in two ways. On the one hand, if spots are not crossed by the transiting planet, then the transit LC is deeper, because the planet occults a larger fraction of the stellar light \citep[see e.g.][]{Czesla2009,Ballerini2012}. The effect is thus that the planet has an apparent radius larger than the true one. On the other hand, as discussed in the introduction of this paper, if spots are crossed during the planetary transit then the transit LC shows localized anomalies.

To test the robustness of \nameofthealgorithm\ in the presence of spots, we first simulate the case of a stellar disk with a spot not occulted by the planet during its transit over the disk. For this purpose, we use the same model of the HAT-P-11 system used above, and we simulate a circular spot 1000~K cooler than the photosphere and radius equal to 0.15R$_*$ (i.e. approximately three times larger than the planet). These parameters depict one of the largest and coolest spots found by \citet{Beky2014} for HAT-P-11. The transit LC is simulated as above, and no random noise is added in order to test the net effect of unocculted starspots on the transit chord reconstruction.

Given this test case, the reconstructed transit chord computed by \nameofthealgorithm\ with $\lambda$=0 is systematically 20~K hotter than the expectations. This is due to the fact that, assuming a spot-free photosphere, \nameofthealgorithm\ overestimates the fraction of stellar light occulted by the planet, and thus the temperature of the occulted stellar photosphere. To fix this bias, we fit the simulated LC using the JKTEBOP code version 34 \citep{Southworth2004}, obtaining an apparent planetary radius of 0.058965 R$_*$, i.e.\ larger than the true radius. We run \nameofthealgorithm\ with this apparent radius, and we find that the simulated transit chord is reconstructed with an accuracy of $\lesssim$3~K, comparable with the accuracy of $\sim$1~K discussed in Sect.~\ref{sec:spotfree}. This indicates that the assumption of a slightly larger planet counteracts the effect of the neglected unocculted photospheric spots, thus improving the accuracy of the algorithm.

This test favors the use of the apparent radius of the planet rather than the true one, thus we advise the user of \nameofthealgorithm\ to fit the transit LC under examination and derive the apparent planetary radius for that given transit. This represents an advantage for \nameofthealgorithm, because the apparent planetary radius is easily measurable from the transit LC, while the true radius is not precisely measurable in case of active photospheres. Nonetheless, we remark that in the most optimistic scenario the LC is affected by a photometric uncertainty as small as 0.1~mmag, which introduces an uncertainty of $\lesssim$30~K in the transit chord reconstruction (Sect.~\ref{sec:spotfree}). Thus, if the true planetary radius is used, then the corresponding bias ($\lesssim$20~K in the simulated case) turns out to be negligible, unless more extremely spotted photospheres have to be taken into account.

\begin{figure*}
\centering
\includegraphics[width=.5\linewidth]{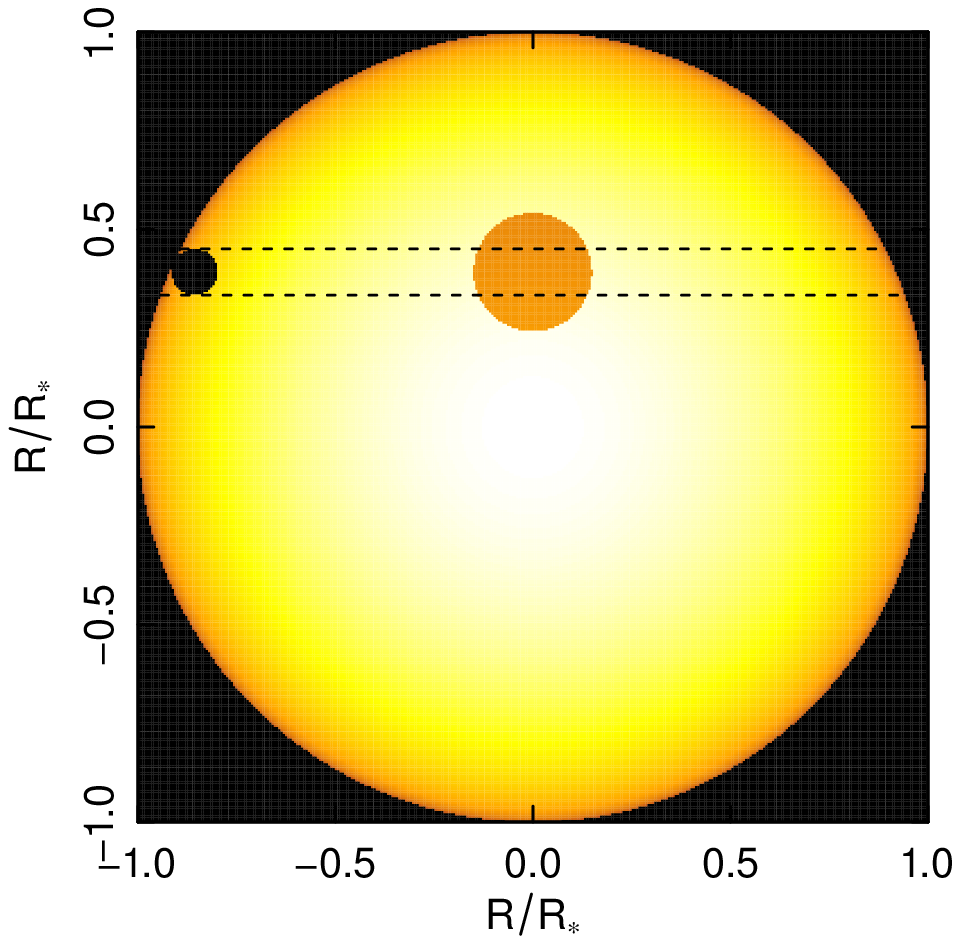}
\includegraphics[width=.4\linewidth,viewport=1 1 425 415,clip]{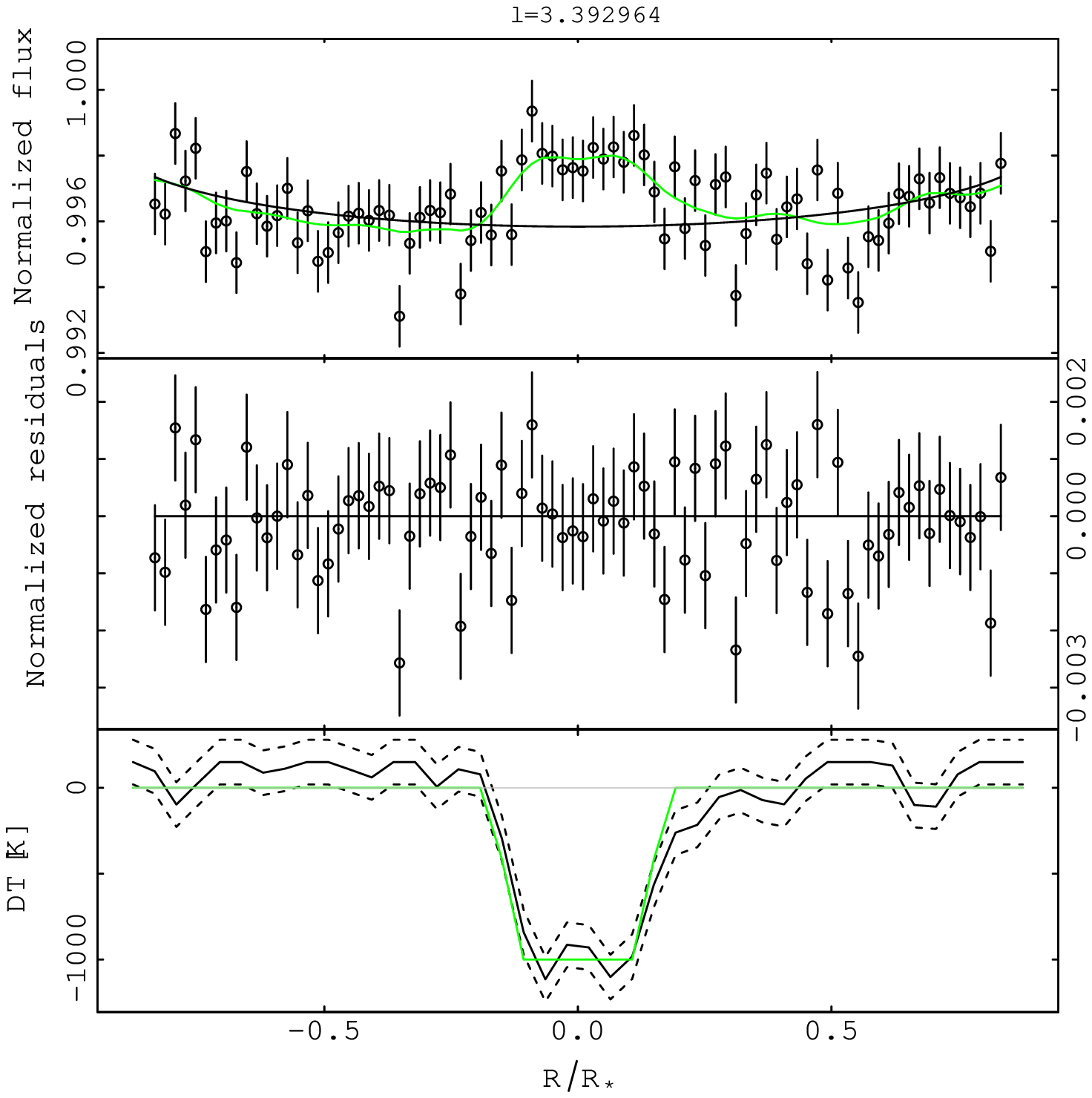}
\caption{\textit{Left panel - }Sketch of the simulated transiting system with a photospheric spot along the transit chord. The horizontal dashed lines delimit the transit chord. Stellar flux increases from orange to white. \textit{Right panel - } Top panel: simulated LC with the unperturbed model and the transit fit represented by the solid black line and the solid green line respectively. Middle: residuals of the transit fit. Bottom: the reconstructed transit chord. In the bottom panel, the uncertainty on the reconstruction is shown in dashes, while the solid green line represents the simulated temperature profile of the chord (see text for better details).}\label{fig:spotted7}
\end{figure*}

Next, we test the robustness of \nameofthealgorithm\ in the presence of transit anomalies given by spot crossing events, which is the main goal of our algorithm. We first simulate the same photospheric spot as above placed at the center of the transit chord (Fig.~\ref{fig:spotted7}, top panel), and we simulate the corresponding light curve, adding a random photometric noise of 1~mmag (Fig.~\ref{fig:spotted7}, middle panel). We fit the LC assuming the apparent planetary radius obtained with JKTEBOP (0.058965~R$_*$, see above), and show the result in the bottom panel of Fig.~\ref{fig:spotted7}: the reconstructed chord obtained with $\lambda=3$  (as returned by the automatic optimization) clearly shows a cooler interval corresponding to the anomaly in the LC, with width and temperature contrast consistent with the simulated spot within the uncertainties. For the sake of comparison, the green solid line in the bottom panel of Fig.~\ref{fig:spotted7} shows the average of the temperature inside the cells drawn over the simulated photosphere.


To compute the uncertainty $\sigma_T$ on the temperature along the transit chord we proceed in the following way. First of all, we simulate the transit LC with the same parameters of the given star and planet, assuming a clear photosphere: the fit of this noise-free LC would provide a noise-free temperature profile of the unspotted transit chord, with the exception of small numerical instabilities. Then, we compute the residuals of the best fit of the observed LC and we add them to the simulated noise-free LC. This gives us a simulated LC with the same photometric accuracy of the observed one. Finally, we run \nameofthealgorithm\ on this noisy simulated LC using the same $\lambda$ optimized for the observed LC, and we compute the standard deviation of the temperature of the grid cells as an estimate of $\sigma_T$. This is an average uncertainty on the reconstructed temperature profile, obtained with the same noise model of the data.

\begin{figure*}
\centering
\includegraphics[width=.5\linewidth]{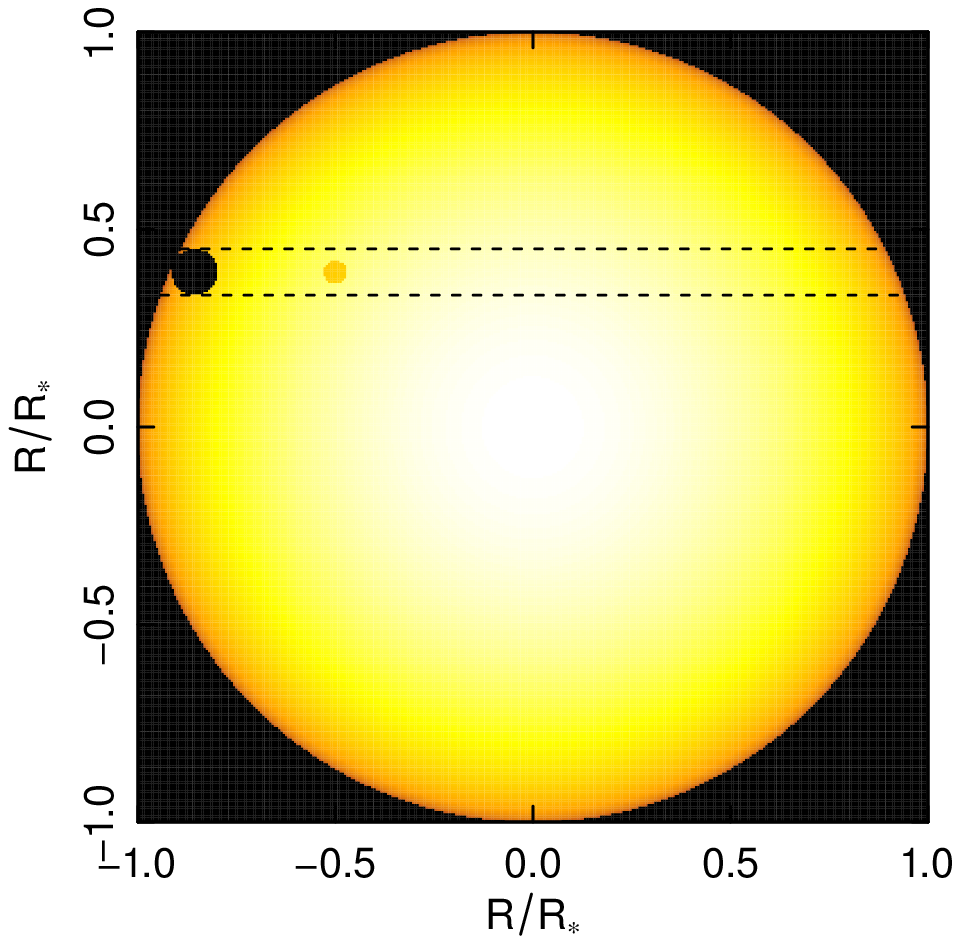}
\includegraphics[width=.4\linewidth,viewport=1 1 425 415,clip]{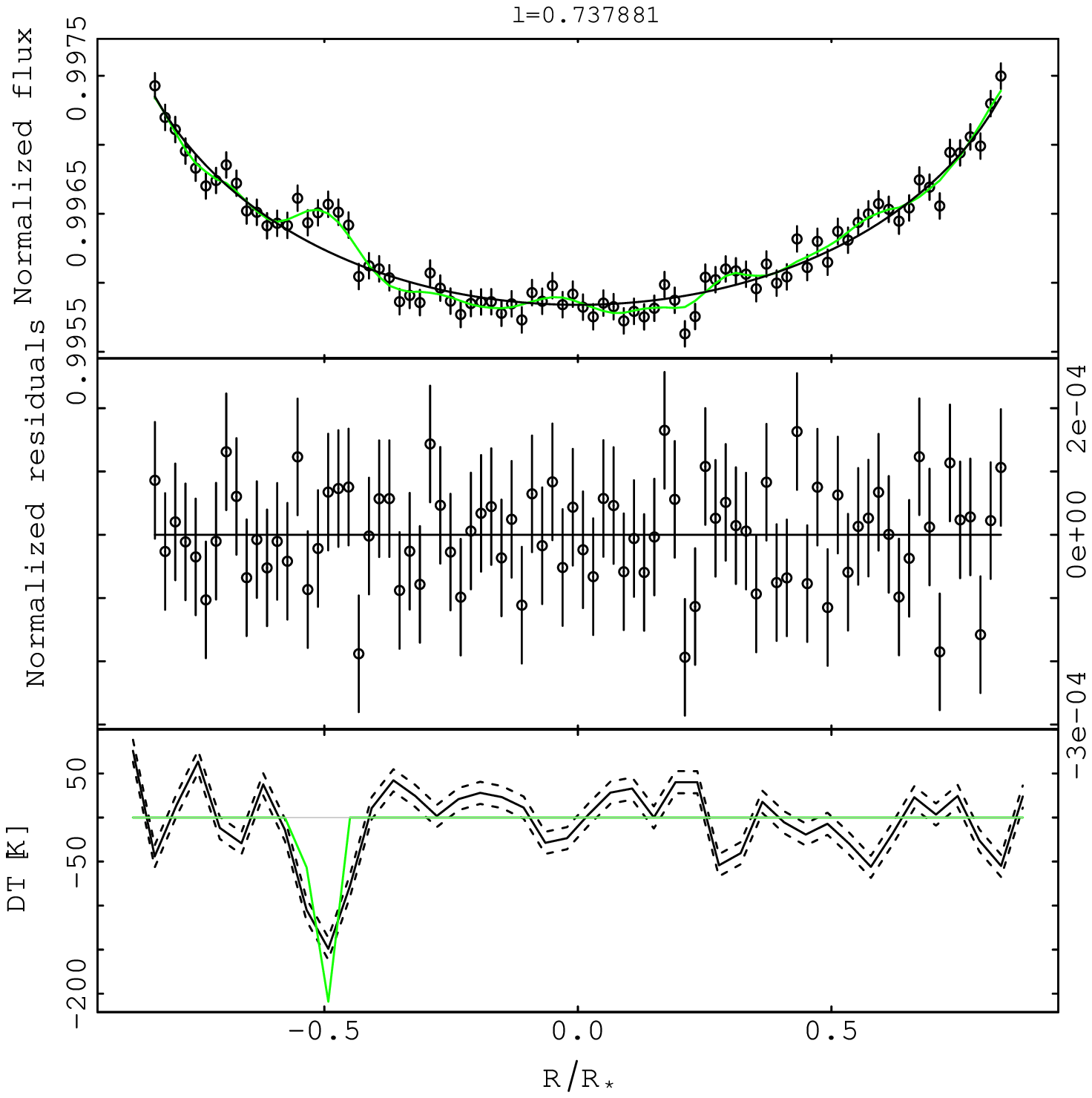}
\caption{Same as in Fig.~\ref{fig:spotted7}, for the test case with a smaller and hotter spot as discussed in the text.}\label{fig:spotted8}
\end{figure*}

Finally, we consider a scenario in which the spot is smaller and hotter, in order to reduce the amplitude of the anomaly with respect to the transit profile depth. In particular, we simulate a circular spot with half the planetary radius and 500~K cooler than the photosphere, placed along the transit chord but off-center, in order to introduce also some asymmetry in the system. Given that this spot has less contrast with respect to the previous test case, its signal is below the 1~mmag photometric uncertainty simulated above. We thus add a smaller photometric noise of 0.1~mmag, which is slightly smaller than the signal expected from the spot crossing event (see right panel in Fig.~\ref{fig:spotted8}). We find that, even if the anomaly is at the limit of the photometric noise, the spot can be detected with a regularization parameter as small as $\lambda=1$. Nonetheless, we remark that the spot-to-photosphere contrast ($\rm \Delta T\gtrsim$-200~K) underestimates the simulated one (-500~K). This is due to two reasons. First of all, the regularization operated by the algorithm tends to smooth out the discontinuities in temperature profile of the chord. Secondly, and most importantly, in this case the spot does not cover the full height of the reconstructed chord, which is equal to the diameter of the planet (see Fig.~\ref{fig:geometry}). The algorithm is not capable to reconstruct the flux distribution perpendicular to the transit chord, then the net result is that along any cell in the transit chord (Fig.~\ref{fig:geometry}) the flux missing due to the cooler temperature of the spot is redistributed over the entire surface of the cell. As a consequence, the temperature of the cells returned by the algorithm is an upper limit for the spot's temperature.

\section{Comparison with previous results on real transits}\label{sec:comparison}

In Sect.~\ref{sec:simulations} we tested \nameofthealgorithm\ against a few simulated cases. In this section we now compare it with other models that have been already discussed in the recent literature. To this purpose, we apply \nameofthealgorithm\ to a few test cases that have been analyzed with similar reconstruction algorithms.

\subsection{CoRoT-2}\label{sec:corot2}

CoRoT-2a (GSC 00465-01282) is an active young G dwarf (\teff=5630~K) hosting the Hot Jupiter CoRoT-2b ($R_p=1.466 R_J$) which transits the stellar disk every $\sim1.74$ days \citep{Alonso2008}. The planetary transit LCs observed with the CoRoT satellite \citep{Baglin2006} show both deformations with respect to the unperturbed transit profile and changes in the transit depth. These features have been attributed to the ubiquitous presence of photospheric dark spots \citep{Lanza2009,SilvaValio2010,SilvaValio2011,Bruno2016}.

\citet{Wolter2009} analyze the most prominent transit anomaly in the whole photometric monitoring of CoRoT-2, found in the transit close to JD=2454335.0, with the aim of modeling the spot crossed by the planet. The authors claim that the LC does not constrain the spot's size, shape and contrast with respect to the photosphere. For this reason they limit their models to circular spots or circle segments enclosed in the transit chord. They also adopt two fixed values for the spot's contrast (i.e. its temperature), consistently with the analysis of \citet{Lanza2009}. The authors conclude that the spot is either $\sim$350~K cooler than the photosphere and is of the same size of the planet, or it is $\sim$1200~K cooler and is as small as half the planet size. They do not discuss the scenario of a large spot centered out of the chord.

\begin{figure}
\centering
\includegraphics[width=\linewidth,viewport=1 1 425 415,clip]{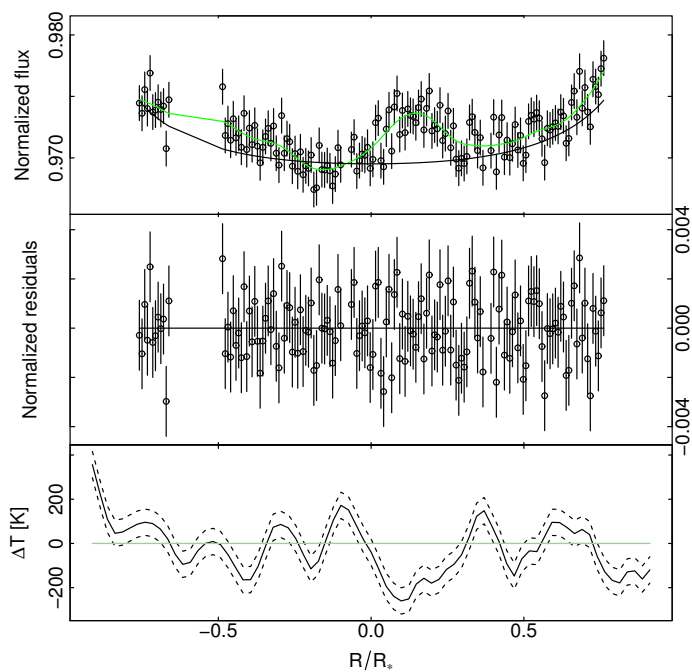}
\caption{\textit{Top panel - } The CoRoT-2 LC discussed in the text. The unperturbed model and the transit fit are represented by the solid black line and the solid green line respectively. \textit{Middle - } Residuals of the transit fit. \textit{Bottom - } The reconstructed transit chord with the uncertainty on the reconstruction shown in dashes.}\label{fig:corot2}
\end{figure}

We analyze the same transit LC discussed by \citet{Wolter2009}. We downloaded the latest reduction of the data from the CoRoT archive, we extracted the time interval around the analyzed transit, we kept the photometric points with good quality as flagged by the reduction pipeline, and we normalized the transit profile with a straight line fitting the LC right before the ingress and after the egress of the planet. The final LC is shown in the top panel of Fig.~\ref{fig:corot2}.

We fit the LC with \nameofthealgorithm\ assuming the parameters in Table~\ref{tab:corot2} and using the photometric passband of CoRoT. We also adopt the apparent planetary radius (see Sect.~\ref{sec:spottedStellarDisk}) $R_p=0.1617~R_*$ as obtained fitting the LC with JKTEBOP after masking out the transit anomaly, in order to fit only the data points not affected by the intersected spot. A posteriori, we find that the Lagrangian multiplier selected by \nameofthealgorithm\ ($\lambda$=1.08) does not give a smooth reconstruction of the transit chord. This is likely an effect of the photometric uncertainty of 1.5 mmag of the LC (Sect.~\ref{sec:spotfree}). We thus artificially set $\lambda$=6.25 and find clear evidence of an area $\sim$300~K cooler than the photosphere slightly after the mid transit and extending over $\simeq0.3 R_*$ in the projected plane.

As discussed in Sect.~\ref{sec:spottedStellarDisk}, we remark that \nameofthealgorithm\ does not provide any information on the flux distribution inside each individual grid cell and that, as a consequence, it provides only an average of the temperature contrast in each cell. Thus we cannot exclude that the cells are only partly covered by the spot, and that the spot's temperature is lower than the one returned by \nameofthealgorithm. Thus we conclude that the spot crossed by CoRoT-2b is $\gtrsim$300~K cooler than the photosphere, consistently with the results of \citet{Wolter2009}.

\begin{table}
\begin{center}
\caption{Parameters of the planetary system CoRoT-2 used in this paper. All data are taken from \citet{Gillon2010}.}\label{tab:corot2}
\begin{tabular}{lr}
\hline\hline
\teff & 5360$\pm$120 K\\
$R_p$ & 0.1658 $R_*$\\
$a$ & 0.02798 AU\\
$i$ & 88.08$^\circ$\\
$e$ & 0.0143\\
$\omega$ & 102$^\circ$\\
$u_1$,$u_2$ & 0.413, 0.293\\
\hline
\end{tabular}

\end{center}
\end{table}

\subsection{HAT-P-11}\label{sec:hatp11}

HAT-P-11a is a V=9.6 K4 dwarf (\teff=4780~K) in the \textit{Kepler} field \citep{Borucki2010}, orbited by a hot Neptune every 4.9 days \citep{Bakos2010}. \citet{Beky2014} aim at analyzing the transits observed by the \textit{Kepler} telescope. To this purpose, they develop \texttt{SPOTROD}, a Monte Carlo algorithm which models the transit anomalies assuming a number of circular photospheric spots. In particular, in their paper they extensively show the results for the transits occurring at BJD=2454967.6 and BJD=2455671.4. In the first case, they find evidence of a spot entirely lying under the transit chord located $\sim0.31 R_*$ ahead the mid-transit, with radius $\simeq0.05 R_*$ and spot-to-photosphere flux ratio of $\simeq0.4$. For the second transit, the authors find two degenerate models for the crossed spot, both with radius $\simeq0.23 R_*$ and flux ratio of $\sim0.8$. The two degenerate models are centered off the transit chord, $\sim0.4 R_*$ after the mid-transit, and they basically differ by the fact that the two circular spots are centered either above or below the chord. Based on the geometry of their model, we derive that the length of the intersection between the spot and the chord in the projected plane is $0.08 R_*$ and $0.3 R_*$ in the two transits respectively.

\begin{figure*}
\centering
\includegraphics[width=.4\linewidth]{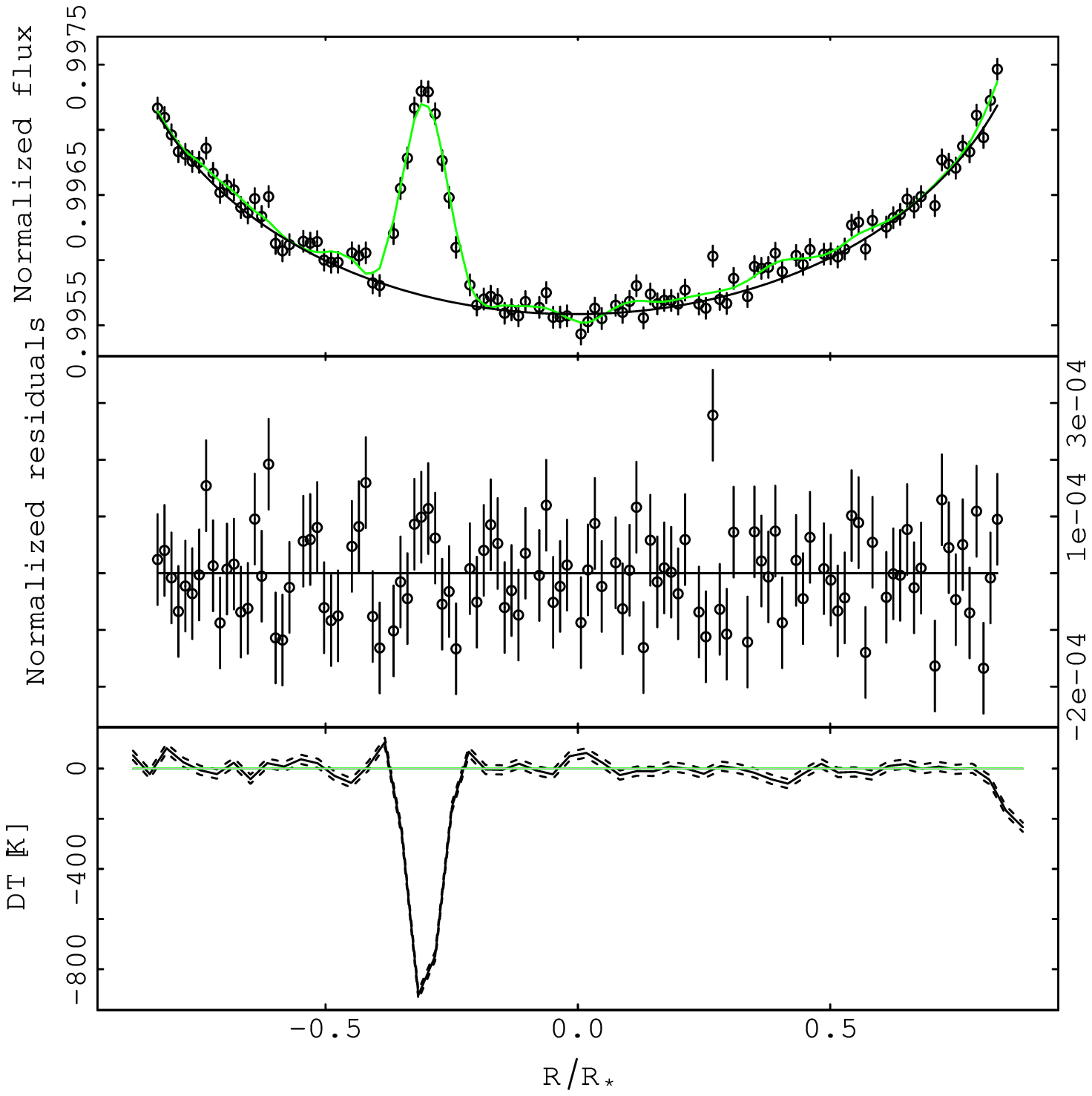}
\hspace{1cm}
\includegraphics[width=.4\linewidth]{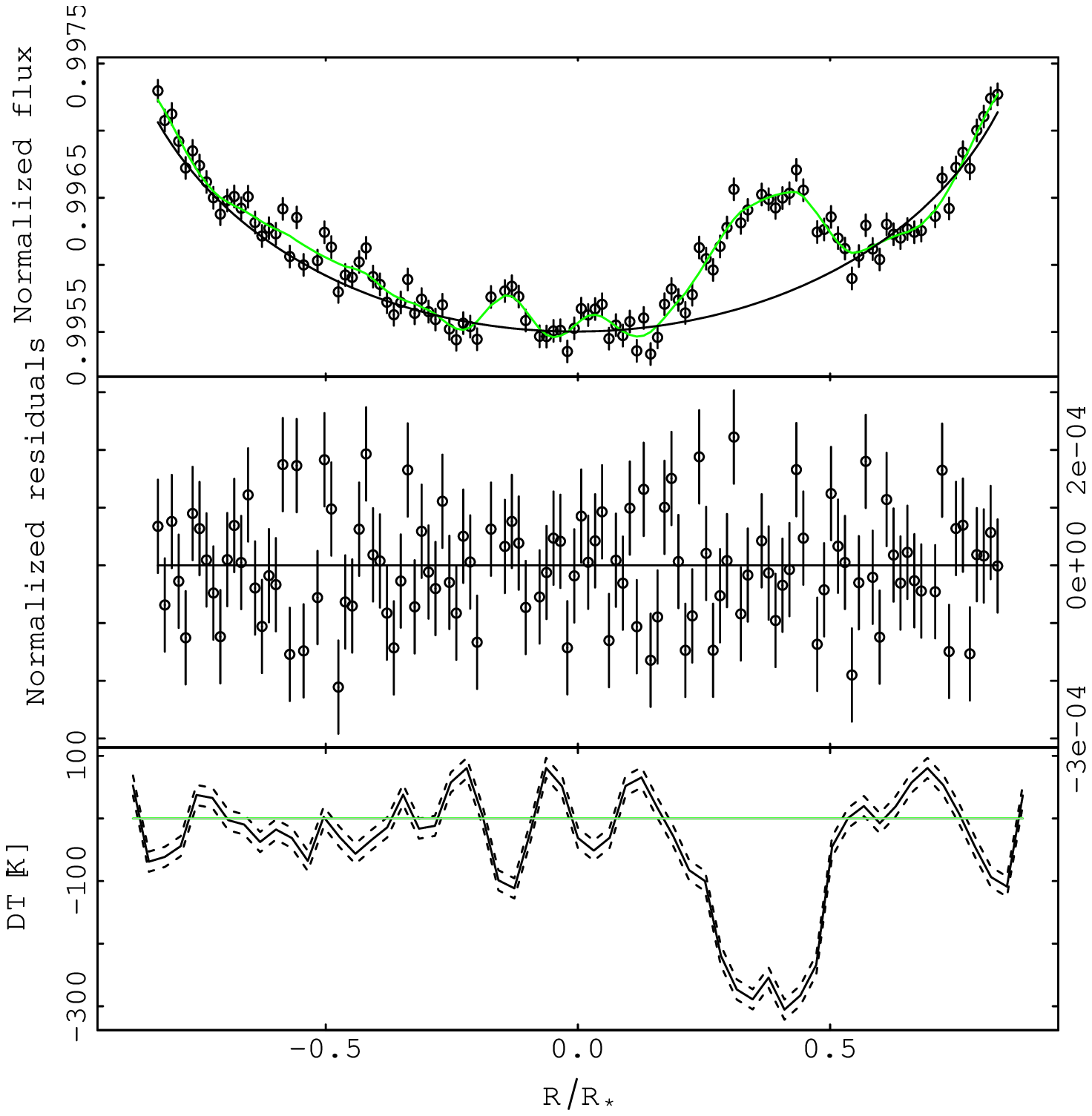}
\caption{Same as in Fig.~\ref{fig:corot2} for the HAT-P-11 system discussed in the text. The left and right panels show the result for the transits occurring on BJD=2454967.6 and BJD=2455671.4 respectively.}\label{fig:hatp11}
\end{figure*}

For the sake of comparison, we run \nameofthealgorithm\ on the same two transit LCs discussed above, using Kepler's passband. The parameters of the planetary system needed as input are listed in Table~\ref{tab:parameters}. The apparent planetary radius obtained with JKTEBOP (see Sect.~\ref{sec:spottedStellarDisk}) is 0.06006 $R_*$ and 0.06063 $R_*$ for the two transits respectively. The output of the algorithm is shown for both cases in Fig.~\ref{fig:hatp11}.

For the first transit \nameofthealgorithm\ optimizes $\lambda=0.34$. The reconstructed anomaly is centered at $\simeq-0.3 R_*$ and is $d\simeq0.15 R_*$ wide. The fitted spot-to-photosphere temperature contrast is $\Delta T\simeq$-900~K, which corresponds to a flux ratio of 0.53.

For the second transit \nameofthealgorithm\ returns $\lambda=0.89$. The reconstructed anomaly is centered at $\simeq0.4 R_*$ and its width is $\simeq0.3 R_*$. The fitted spot-to-photosphere temperature contrast is $\Delta T\simeq$-350~K, which leads to a flux ratio of 0.8.

\begin{figure*}
\centering
\includegraphics[width=.4\linewidth]{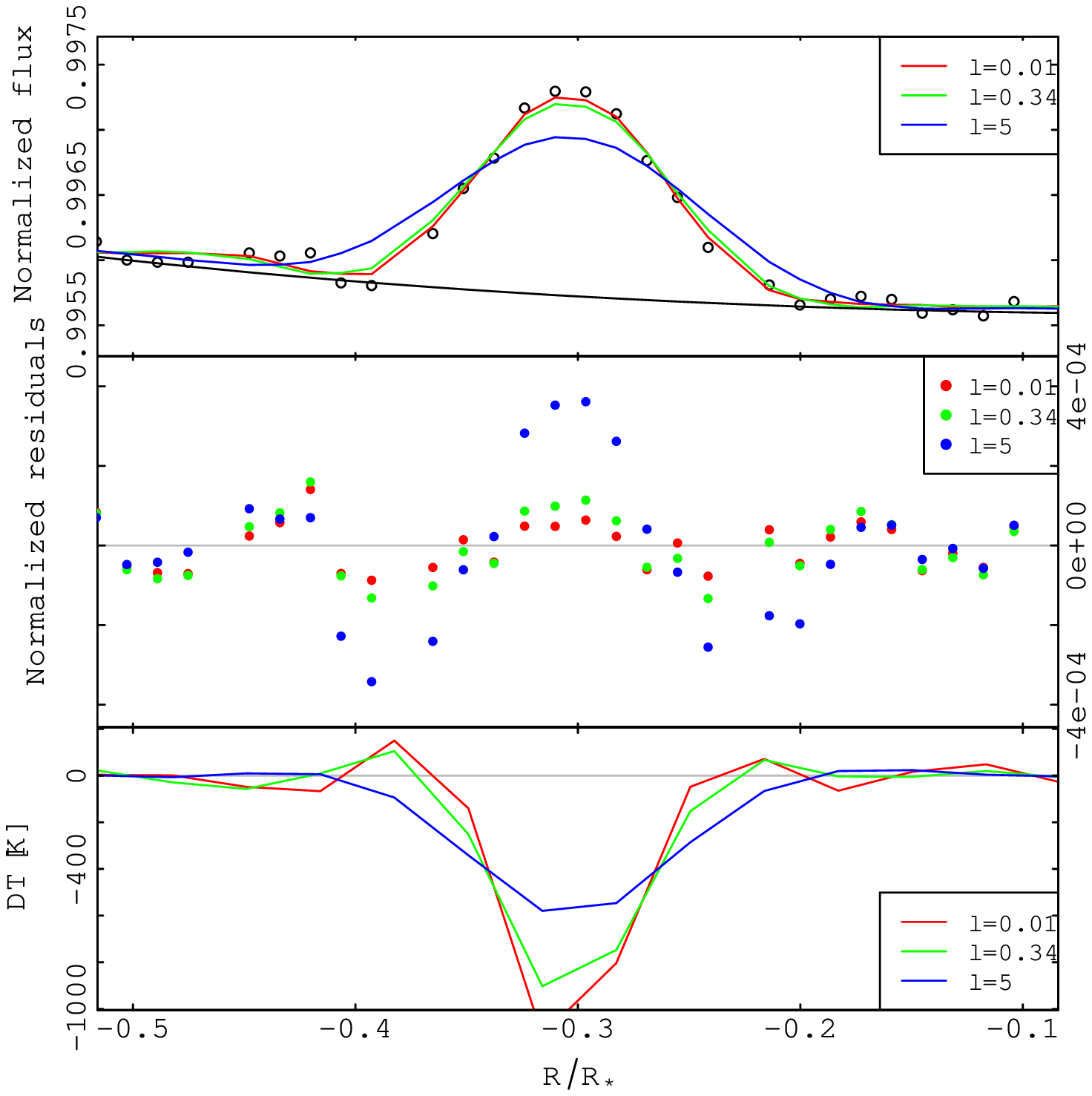}
\includegraphics[width=.4\linewidth]{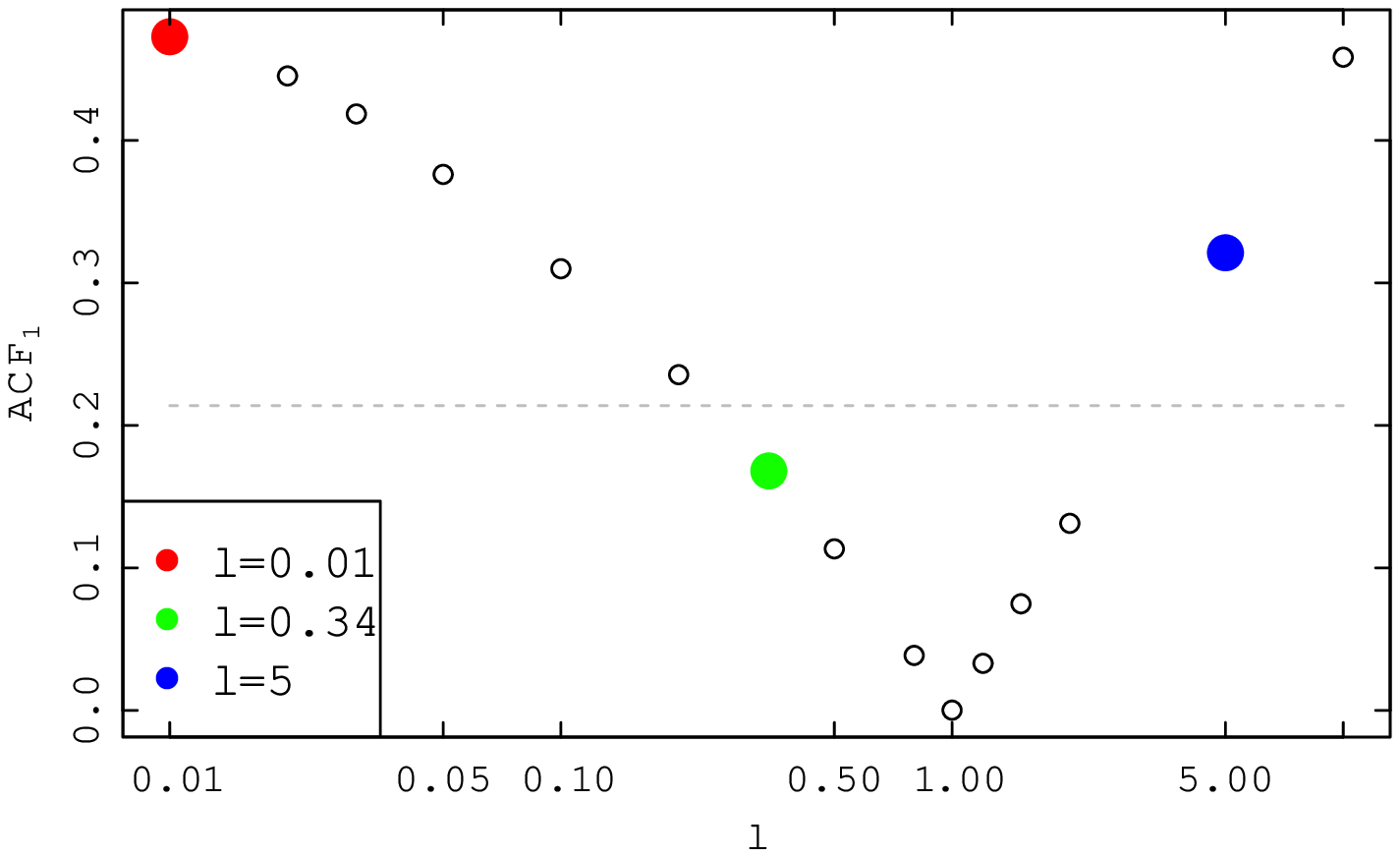}
\caption{\textit{Left panel - }Zoom in of the left panel in Fig.~\ref{fig:hatp11}, showing the effect of different values of $\lambda$ in the reconstruction algorithm. \textit{Right panel - }Autocorrelation $ACF_1$ between neighboring cells in the residuals vs. $\lambda$. The three examples shown in the left panel are marked in colors, while the gray dashes represent the $1.96/\sqrt{N}$ threshold adopted in Sect.~\ref{sec:algorithm} (see text for discussion).}\label{fig:lambdaramp}
\end{figure*}

In both cases our results are thus generally consistent with what is obtained using \texttt{SPOTROD}. Only for the first transit we find some discrepancy in the spot-to-photosphere flux ratio. A closer look to the best fit of the LC returned by \nameofthealgorithm\ reveals that the algorithm systematically underestimates the observed flux at the center of the transit anomaly by more than the photometric uncertainty. This is due to the smoothing of the best-fit model introduced by the regularizing constraints in the algorithm. As a matter of fact, the best fit returned by \nameofthealgorithm\ follows more closely the data if we artificially reduce $\lambda$ down to 0.01. Correspondingly, the temperature contrast decreases to $\Delta T\simeq-1200$ K, which corresponds to a flux ratio of $\sim$0.38, consistent with the results of \citet{Beky2014}. Nonetheless, the residuals show some degree of autocorrelation, introduced by the inversion algorithm (see Sect.~\ref{sec:spotfree}). The purpose of the automatic optimization discussed in Sect.~\ref{sec:algorithm} is to find the smallest Lagrangian multiplier which returns non autocorrelated residuals. In this particular case, the best Lagrangian multiplier is $\lambda$=0.34. If, on the other hand, we artificially use large multipliers, the best fit tend to be over-smoothed and, correspondingly, the residuals get more autocorrelated as $\lambda$ increases. In the left panel of Fig.~\ref{fig:lambdaramp} we plot the best fit obtained with three different values of $\lambda$, while in the right panel we plot the autocorrelation of the residuals computed at different values of $\lambda$.

\subsection{HAT-P-36}\label{sec:hatp36}

HAT-P-36a is a V=12.5 G5 dwarf (\teff=5620~K) orbited by a hot Jupiter ($R_p=1.3 R_J$) approximately every 1.3 days \citep{Bakos2012}. In their detailed study of the system, \citet{Mancini2015} analyzed the anomalies in the transit LCs of the system. They find the most evident anomalies in the transit LC in the Gunn $r$ band, collected on April,15th 2013 with the BFOSC imager mounted on the 1.52~m Cassini Telescope at the Astronomical Observatory of Bologna in Loiano, Italy. They use the GEMC+PRISM codes \citep{Tregloan2013,Tregloan2015}, which basically combines Monte Carlo and genetic algorithms to simulate circular spots on the stellar surface. In their analysis of the aforementioned LC, they find evidence of two spots (one after the second contact and the other before the third contact) with T$\simeq$5000~K and radius $\simeq0.2 R_*$. In particular, the planet disk does not enter completely into the spots or, in other words, the spots do not completely cover the occulted belt along the vertical direction. Given the geometrical model used by the GEMC+PRISM algorithm, we derive that the length of the intersection of the chord with each of the stellar spots is $\lesssim0.16 R_*$ in the projected plane.

\begin{figure}
\centering
\includegraphics[width=\linewidth,viewport=1 1 425 415,clip]{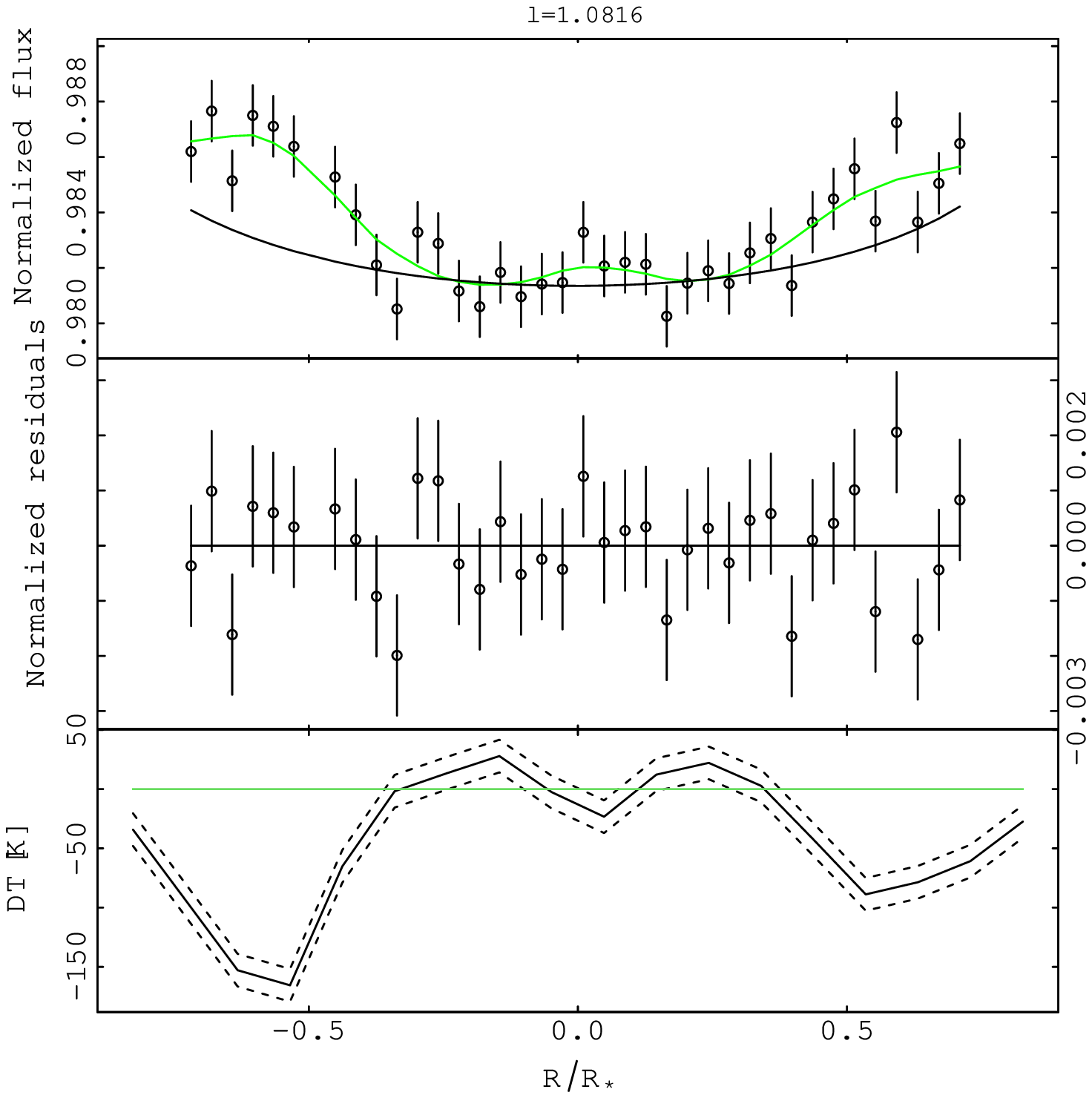}
\caption{Same as in Fig.~\ref{fig:corot2} for the transit LCs of the HAT-P-36 system discussed in the text.}\label{fig:hatp36}
\end{figure}

We analyze the normalized LC provided by \citet{Mancini2015}, using \nameofthealgorithm\ with the parameters listed in Table~\ref{tab:hatp36}. We do not list the argument of periastron $\omega$ as it is meaningless in case of circular orbits. We also use the apparent radius $R_p=0.1273 R_*$ obtained fitting the LC with JKTEBOP. The output of the transit chord reconstruction, obtained with regularization $\lambda=1.08$, is shown in Fig.~\ref{fig:hatp36}. We find that the two spot crossing events occur at the borders of the reconstructed chord. Nonetheless, our fit suggests that the two sections of the projected transit chord intersected by the spots are $\gtrsim0.12 R_*$, while the temperature contrast is between $\Delta T$=-100~K and $\Delta T=$-200~K, consistently different from what \citet{Mancini2015} found ($\Delta T\simeq$-600~K). This may be due to the fact that, according to \citet{Mancini2015}, the two spots do not intersect the whole height of the occulted belt. As we discussed in Sect.~\ref{sec:spottedStellarDisk}, when the grid cells are only partly covered by the spot, the temperature returned by \nameofthealgorithm\ is an intermediate value between the \teff\ and the true temperature of the spot $T_s$. Let's assume that the fitted flux $F$ is given by the weighted mean
\begin{equation}
F=\left(1-\alpha\right)F_*+\alpha F_s,
\end{equation}
where $\alpha$ is the fraction of the grid cells covered by the spot, while $F_*$ and $F_s$ are the temperature of the quiet photosphere and the spot respectively. After a few math, we obtain
\begin{equation}
\frac{F_s}{F_*}=1-\frac{1-F/F_*}{\alpha},
\end{equation}
where $\frac{F_s}{F_*}$ is the spot-to-photosphere flux contrast and $\frac{F}{F_*}\simeq$0.8 is the contrast corresponding to the $\Delta T\simeq$150~K returned by \nameofthealgorithm. Following the scenario depicted by \citet{Mancini2015} in their Fig.~8, if we assume $\alpha\lesssim0.5$ then we obtain $\frac{F_s}{F_*}\lesssim$0.6, which corresponds to a temperature contrast of $\lesssim$-400~K, closer to their result.

\begin{table}
\begin{center}
\caption{Parameters of the planetary system HAT-P-36 used in this paper. All data are taken from \citet{Mancini2015}.}\label{tab:hatp36}
\begin{tabular}{lr}
\hline\hline
\teff & 5620$\pm$40 K\\
$R_p$ & 0.1243 $R_*$\\
$a$ & 0.02388 AU\\
$i$ & 85.86$^\circ$\\
$e$ & 0\\
$u_1$,$u_2$ & 0.24, 0.40\\
\hline
\end{tabular}

\end{center}
\end{table}

\subsection{GJ1214}

GJ1214a is a M4.5 red dwarf orbited by a super-Earth ($R_p\simeq2.6R_\oplus$) every 1.58 days \citep{Charbonneau2009}. On May,17th 2012, \citet{Nascimbeni2015} observed a planetary transit simultaneously in the B and R photometric bands using the LBC camera mounted at the double 8.4m Large Binocular Telescope. The photometric uncertainty of the observed LCs is $\lesssim$1~mmag. Both transit LCs show a clear signature of the crossing of a spot near the egress of the planet. Analyzing the shape (width and height) of the transit anomaly in the two bands, the authors conclude that the projected spot is as large as the planet, and is $\sim$110~K cooler than the photosphere.

\citet{Nascimbeni2015} also analyze the residuals with respect to the best fit, dividing the LCs in three non overlapping sections: the off-transit, the spot (corresponding to the transit anomaly) and the in-transit (which does not include the anomaly). They find no correlation between the B and R off-transit residuals, which is indicative of the fact that no correlated noise (due to e.g. instrumental or atmospheric effects) affects the data. Conversely, in the transit anomaly the B and R residuals show a clear correlation, as it is expected when a large spot is occulted. Most interestingly, the same correlation is present in the in-transit residuals, suggesting that the whole transit chord is characterized by inhomogeneities.

\begin{figure*}
\centering
\includegraphics[width=.4\linewidth,viewport=1 1 425 415,clip]{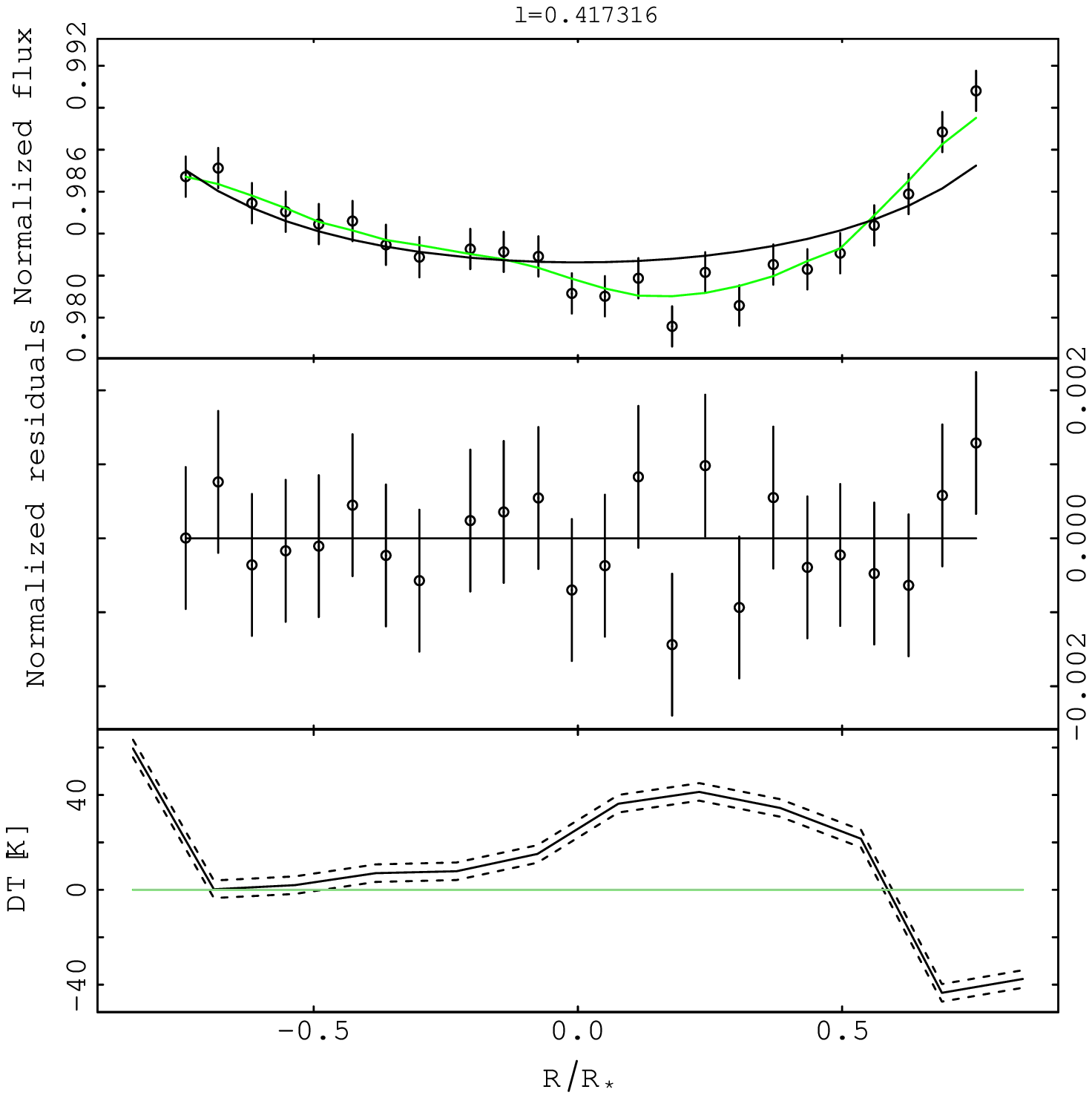}
\hspace{1cm}
\includegraphics[width=.4\linewidth,viewport=1 1 425 415,clip]{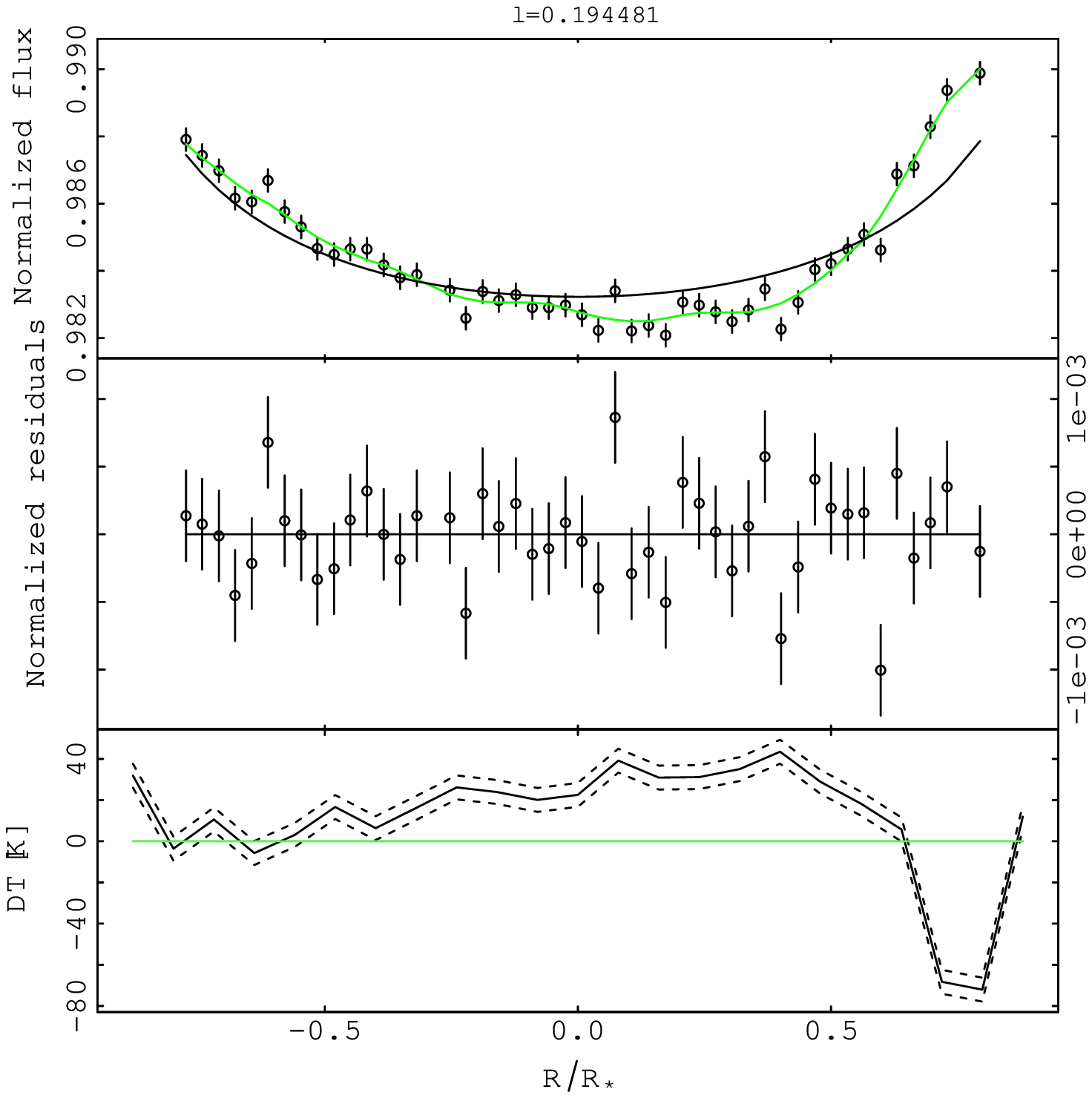}
\caption{Same as in Fig.~\ref{fig:corot2} for the transit LCs of the GJ1214 system discussed in the text. The left panel shows the fit of the B band LC, obtained with regularization parameter $\lambda$=0.42. The right panel shows the fit of the R band LC, obtained with regularization parameter $\lambda$=0.19.}\label{fig:gj1214}
\end{figure*}

We use \nameofthealgorithm\ to analyze in further details the same B and R LCs discussed above. In Table~\ref{tab:gj1214} we list the parameters injected in the algorithm. As in Sect.~\ref{sec:hatp36}, we do not list the argument of periastron $\omega$ as it is meaningless in case of circular orbits. We also use the apparent planetary radius derived by \citet{Nascimbeni2015}, i.e.\ 0.1186 R$_*$ and 0.1178 R$_*$ in the B and R bands respectively. The output of the reconstruction is shown in Fig.~\ref{fig:gj1214}. Unfortunately, the spot crossing event occurs too close to the egress and \nameofthealgorithm\ is not able to entirely model it. Nonetheless, both reconstructed transit chords clearly show a photospheric spot $\lesssim$80~K cooler than the photosphere near the egress.

\begin{table}
\begin{center}
\caption{Parameters of the planetary system GJ1214 used in this paper}\label{tab:gj1214}
\begin{tabular}{lrl}
 \hline\hline
 &   & Reference\\
\hline
\teff & 3252$\pm$20 K & \citet{Anglada2013}\\
$R_p$ & 0.1161 R$_*$ & \citet{Carter2011}\\
$a$ & 14.97 R$_*$ & \citet{Bean2010}\\
$i$ & 88.94$^\circ$ & \citet{Bean2010}\\
$e$ & 0 & \citet{Carter2011}\\
$u_{1,B}$,$u_{2,B}$ & 0.5259, 0.3296 & \citet{Nascimbeni2015}\\
$u_{1,R}$,$u_{2,R}$ & 0.4326, 0.3323 & \citet{Nascimbeni2015}\\
\hline
\end{tabular}
\end{center}
\end{table}

The transit LCs also show that after the mid-transit and before the anomaly the transit depth increases by $\sim$0.2\% in the B band and $\sim$0.1\% in the R band. This suggests that the part of the stellar disk occulted by the planet after mid-transit is brighter than the average stellar disk. Accordingly, the temperature contrast returned by \nameofthealgorithm\ suggests that the second half of the transit chord is hotter by $\sim$50~K than the quiet photosphere. The increase in brightness may be due to a large group of photospheric faculae, extending for $\sim$0.5~R$_*$ and neighboring the spot detected near the stellar limb. This scenario is consistent with the solar case, where spots are often accompanied by faculae \citep{Wiegelmann2014}. Moreover, we remark that the brightening effect seems to increase approaching the stellar limb, as indicated by \citet{Foukal1991,Solanki2006}, and that it is more pronounced in the B band than in the R band, consistently with the fact that the facular contrast increases toward shorter wavelengths \citep{Chapman1977,Ermolli2006}.

We remark that \nameofthealgorithm\ is only suited for modeling cold spots on the stellar surface, i.e. it is not capable to model warm regions such as faculae or plages, for which a proper spectrum should be taken into account. Even though the numerical value of the temperature contrast is unreliable, still the best fit of the LCs clearly indicates an extended warm section of the transit chord, consistently between the B and R data sets. The presence of this region supports the hypothesis of \citet{Nascimbeni2015}, which postulated the presence of additional inhomogeneities on the stellar surface to explain the correlated noise in the in-transit residuals. The presence of faculae in the photosphere of GJ1214a has also been recently postulated by \citet{Rackham2017}.

Another interpretation of the analyzed transit LC is that cool spots are ubiquitous along the transit chord. In this scenario, even masking out the most evident transit anomaly, the lower envelope of the LC is shallower than the LC expected in the case of a quiet transit chord, i.e. the apparent planetary radius is underestimated (see Sect.~\ref{sec:spottedStellarDisk}). If we arbitrarily increase the apparent $R_p$ used to reconstruct the transit chord by 5\%, then the $\Delta T$ is smaller (i.e. the transit chord is cooler) by $~\sim$20~K, i.e.\ there are no faculae in the transit chord and the temperature contrast is of the order of $\sim$-100~K, as found by \citet{Nascimbeni2015}.

\section{Conclusions}

In this paper we present \nameofthealgorithm, a new fast algorithm which analyzes the anomalies in the transit LCs with the aim of reconstructing the flux distribution, and the temperature profile, of the transit chord. The only inputs needed by the code are the parameters of the planetary system, the LD coefficients and the model spectrum for unspotted and spotted photosphere. The presence of unocculted spots, not reconstructed by \nameofthealgorithm, can be accounted for using the apparent radius of the planet in place of the true radius.

We test the performances of \nameofthealgorithm\ using simulated transits. We show that \nameofthealgorithm\ is generally able to reconstruct the crossed spots for a photometric accuracy better than 1~mmag, and that the temperature contrast is returned with an uncertainty of $\lesssim$100~K. We also test \nameofthealgorithm\ with real transits available in literature which have been already analyzed with alternative codes. These comparisons show that \nameofthealgorithm\ is consistent with previous approaches, also in the case of more sophisticated algorithms.

\nameofthealgorithm\ is available as a web interface at the URL \url{www.oact.inaf.it/tosc}, where the user can run the algorithm feeding a few input files and retrieving the output. The web page contains also extensive instructions and some example data. The source code is available for download too.

\begin{acknowledgements}
G.S., V.N. and I.P. acknowledge financial support from \lq\lq\ Accordo ASI--INAF\rq\rq\ n. 2013-016-R.0 July 9, 2013.

G.S.\ dedicates this paper to his alive and kicking father, after whom \nameofthealgorithm\ is named.

\end{acknowledgements}


\begin{thebibliography}{}
\bibitem[Alonso et al.(2008)]{Alonso2008} Alonso, R., Auvergne, M., Baglin, A., et al.\ 2008, \aap, 482, L21 
\bibitem[Andersen \& Korhonen(2015)]{Andersen2015} Andersen, J.~M., \& Korhonen, H.\ 2015, \mnras, 448, 3053 
\bibitem[Anglada-Escud{\'e} et al.(2013)]{Anglada2013} Anglada-Escud{\'e}, G., Rojas-Ayala, B., Boss, A.~P., Weinberger, A.~J., \& Lloyd, J.~P.\ 2013, \aap, 551, A48 
\bibitem[Ballerini et al.(2012)]{Ballerini2012} Ballerini, P., Micela, G., Lanza, A.~F., \& Pagano, I.\ 2012, \aap, 539, A140 
\bibitem[Baglin et al.(2006)]{Baglin2006} Baglin, A., Auvergne, M., Boisnard, L., et al.\ 2006, 36th COSPAR Scientific Assembly, 36, 3749 
\bibitem[Bakos et al.(2010)]{Bakos2010} Bakos, G.~{\'A}., Torres, G., P{\'a}l, A., et al.\ 2010, \apj, 710, 1724 
\bibitem[Bakos et al.(2012)]{Bakos2012} Bakos, G.~{\'A}., Hartman, J.~D., Torres, G., et al.\ 2012, \aj, 144, 19 
\bibitem[Baraffe et al.(2015)]{Baraffe2015} Baraffe, I., Homeier, D., Allard, F., \& Chabrier, G.\ 2015, \aap, 577, A42 
\bibitem[Bean et al.(2010)]{Bean2010} Bean, J.~L., Miller-Ricci Kempton, E., \& Homeier, D.\ 2010, \nat, 468, 669 
\bibitem[B{\'e}ky et al.(2014a)]{Beky2014a} B{\'e}ky, B., Holman, M.~J., Kipping, D.~M., \& Noyes, R.~W.\ 2014, \apj, 788, 1 
\bibitem[B{\'e}ky et al.(2014b)]{Beky2014} B{\'e}ky, B., Kipping, D.~M., \& Holman, M.~J.\ 2014, \mnras, 442, 3686 
\bibitem[Berdyugina(2005)]{Berdyugina2005} Berdyugina, S.~V.\ 2005, Living Reviews in Solar Physics, 2, 8 
\bibitem[Borsa et al.(2015)]{Borsa2015} Borsa, F., Scandariato, G., Rainer, M., et al.\ 2015, \aap, 578, A64 
\bibitem[Borucki et al.(2010)]{Borucki2010} Borucki, W.~J., Koch, D., Basri, G., et al.\ 2010, Science, 327, 977 
\bibitem[Bruno et al.(2016)]{Bruno2016} Bruno, G., Deleuil, M., Almenara, J.-M., et al.\ 2016, \aap, 595, A89 
\bibitem[Carter et al.(2011)]{Carter2011} Carter, J.~A., Winn, J.~N., Holman, M.~J., et al.\ 2011, \apj, 730, 82 
\bibitem[Chapman \& McGuire(1977)]{Chapman1977} Chapman, G.~A., \& McGuire, T.~E.\ 1977, \apj, 217, 657 
\bibitem[Charbonneau et al.(2009)]{Charbonneau2009} Charbonneau, D., Berta, Z.~K., Irwin, J., et al.\ 2009, \nat, 462, 891 
\bibitem[Chatfield(1980)]{Chatfield1980} Chatfield, C.\ 1980, The Analysis of Time Series: An Introduction, Chapman and Hall ed.
\bibitem[Claret(2004)]{Claret2004} Claret, A.\ 2004, \aap, 428, 1001 
\bibitem[Czesla et al.(2009)]{Czesla2009} Czesla, S., Huber, K.~F., Wolter, U., Schr{\"o}ter, S., \& Schmitt, J.~H.~M.~M.\ 2009, \aap, 505, 1277
\bibitem[D{\'e}sert et al.(2011)]{Desert2011} D{\'e}sert, J.-M., Charbonneau, D., Demory, B.-O., et al.\ 2011, \apjs, 197, 14 
\bibitem[Ermolli et al.(2007)]{Ermolli2006} Ermolli, I., Criscuoli, S., Centrone, M., Giorgi, F., \& Penza, V.\ 2007, \aap, 465, 305 
\bibitem[Foukal et al.(1991)]{Foukal1991} Foukal, P., Harvey, K., \& Hill, F.\ 1991, \apjl, 383, L89 
\bibitem[Gillon et al.(2010)]{Gillon2010} Gillon, M., Lanotte, A.~A., Barman, T., et al.\ 2010, \aap, 511, A3 
\bibitem[Gray(1992)]{Gray1992} Gray, D.~F.\ 1992, Camb.~Astrophys.~Ser., Vol.~20,,
\bibitem[Houdebine et al.(1995)]{Houdebine1995} Houdebine, E.~R., Doyle, J.~G., \& Koscielecki, M.\ 1995, \aap, 294, 773 
\bibitem[Huber et al.(2010)]{Huber2010} Huber, K.~F., Czesla, S., Wolter, U., \& Schmitt, J.~H.~M.~M.\ 2010, \aap, 514, A39 
\bibitem[Lanza et al.(2009)]{Lanza2009} Lanza, A.~F., Pagano, I., Leto, G., et al.\ 2009, \aap, 493, 193 
\bibitem[Lanza(2016)]{Lanza2016} Lanza, A.~F.\ 2016, Lecture Notes in Physics, Berlin Springer Verlag, 914, 43 
\bibitem[Nascimbeni et al.(2015)]{Nascimbeni2015} Nascimbeni, V., Mallonn, M., Scandariato, G., et al.\ 2015, \aap, 579, A113 
\bibitem[Maldonado et al.(2015)]{Maldonado2015} Maldonado, J., Affer, L., Micela, G., et al.\ 2015, \aap, 577, A132 
\bibitem[Mancini et al.(2015)]{Mancini2015} Mancini, L., Esposito, M., Covino, E., et al.\ 2015, \aap, 579, A136 
\bibitem[Nutzman et al.(2011)]{Nutzman2011} Nutzman, P.~A., Fabrycky, D.~C., \& Fortney, J.~J.\ 2011, \apjl, 740, L10 
\bibitem[Piskunov et al.(1990)]{Piskunov1990} Piskunov, N.~E., Tuominen, I., \& Vilhu, O.\ 1990, \aap, 230, 363 
\bibitem[Pont et al.(2007)]{Pont2007} Pont, F., Gilliland, R.~L., Moutou, C., et al.\ 2007, \aap, 476, 1347 
\bibitem[Press et al.(1992)]{Press1992} Press, W.~H., Teukolsky, S.~A., Vetterling, W.~T., \& Flannery, B.~P.\ 1992, Cambridge: University Press, |c1992, 2nd ed.,  
\bibitem[R core team(2016)]{R2016} R Core Team (2016). R: A language and environment for statistical computing. R Foundation for Statistical Computing, Vienna, Austria. URL https://www.R-project.org/.
\bibitem[Rabus et al.(2009)]{Rabus2009} Rabus, M., Alonso, R., Belmonte, J.~A., et al.\ 2009, \aap, 494, 391 
\bibitem[Rackham et al.(2017)]{Rackham2017} Rackham, B., Espinoza, N., Apai, D., et al.\ 2017, \apj, 834, 151 
\bibitem[Sanchis-Ojeda \& Winn(2011a)]{Sanchis2011} Sanchis-Ojeda, R., \& Winn, J.~N.\ 2011, \apj, 743, 61 
\bibitem[Sanchis-Ojeda et al.(2011b)]{Sanchis2011b} Sanchis-Ojeda, R., Winn, J.~N., Holman, M.~J., et al.\ 2011, \apj, 733, 127 
\bibitem[Scandariato et al.(2016)]{Scandariato2016} Scandariato, G. et al.\ 2017, \lq\lq Frontier Research in Astrophysics - II\rq\rq, F. Giovannelli \& L. Sabau-Graziati (Eds.), PoS-SISSA (submitted)
\bibitem[Schneider(2000)]{Schneider2000} Schneider, J.\ 2000, From Giant Planets to Cool Stars, 212, 284 
\bibitem[Silva(2003)]{Silva2003} Silva, A.~V.~R.\ 2003, \apjl, 585, L147 
\bibitem[Silva-Valio(2008)]{Silva2008} Silva-Valio, A.\ 2008, \apjl, 683, L179 
\bibitem[Silva-Valio et al.(2010)]{SilvaValio2010} Silva-Valio, A., Lanza, A.~F., Alonso, R., \& Barge, P.\ 2010, \aap, 510, A25 
\bibitem[Silva-Valio \& Lanza(2011)]{SilvaValio2011} Silva-Valio, A., \& Lanza, A.~F.\ 2011, \aap, 529, A36 
\bibitem[Soetaert et al.(2009)]{Soetaert2009} Soetaert K., Van den Meersche, K., van Oevelen, D. (2009). limSolve: Solving Linear Inverse Models. R-package version 1.5.1
\bibitem[Solanki et al.(2006)]{Solanki2006} Solanki, S.~K., Inhester, B., \& Sch{\"u}ssler, M.\ 2006, Reports on Progress in Physics, 69, 563 
\bibitem[Southworth et al.(2004)]{Southworth2004} Southworth, J., Maxted, P.~F.~L., \& Smalley, B.\ 2004, \mnras, 351, 1277 
\bibitem[Tregloan-Reed et al.(2013)]{Tregloan2013} Tregloan-Reed, J., Southworth, J., \& Tappert, C.\ 2013, \mnras, 428, 3671 
\bibitem[Tregloan-Reed et al.(2015)]{Tregloan2015} Tregloan-Reed, J., Southworth, J., Burgdorf, M., et al.\ 2015, \mnras, 450, 1760 
\bibitem[Wiegelmann et al.(2014)]{Wiegelmann2014} Wiegelmann, T., Thalmann, J.~K., \& Solanki, S.~K.\ 2014, \aapr, 22, 78 
\bibitem[Winn(2010a)]{Winn2010} Winn, J.~N.\ 2010, Exoplanets, 55 
\bibitem[Winn et al.(2010b)]{Winn2010b} Winn, J.~N., Johnson, J.~A., Howard, A.~W., et al.\ 2010, \apjl, 723, L223 
\bibitem[Wolter et al.(2009)]{Wolter2009} Wolter, U., Schmitt, J.~H.~M.~M., Huber, K.~F., et al.\ 2009, \aap, 504, 561 






\end{thebibliography}
\end{document}